\documentclass{article}
\usepackage[utf8]{inputenc}
\usepackage{geometry}
\usepackage{graphicx} 
\graphicspath{{figures/}}
\usepackage{braket}
\usepackage{amsmath,amssymb}
\usepackage{booktabs}  
\usepackage{placeins}  
\usepackage[font=small,labelfont=bf,skip=6pt]{caption}  
\usepackage{todonotes}
\usepackage{eso-pic} 
\usepackage[numbers,sort&compress]{natbib}
\usepackage{hyperref}  

\newsavebox{\logo}
\savebox{\logo}{\includegraphics[width=1.4in]{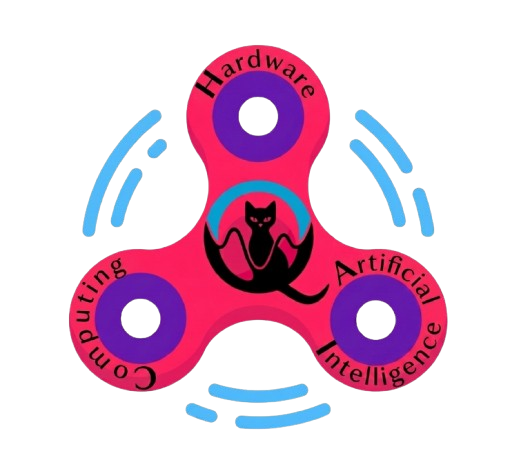}}%
\AddToShipoutPictureFG*{%
  \put(
    \LenToUnit{\dimexpr\paperwidth-1.8in},
    \LenToUnit{\dimexpr\paperheight-1.65in}  
  ){\usebox\logo}%
}

\title{LatentRL-QAS: Transfer via \textbf{L}atent \textbf{R}einforcement \textbf{L}earning for \textbf{Q}uantum \textbf{A}rchitecture \textbf{S}earch}
\author{Maria Ilektra Plakitsi\thanks{Corresponding author: \texttt{m.i.plakitsi@student.tudelft.nl}}, Sebastian Feld, Akash Kundu\\[0.6em]
    \small QuTech and Faculty of Electrical Engineering, Mathematics and Computer Science,\\
    \small Delft University of Technology, Delft, The Netherlands}
\date{}

\newcommand{\method}{LatentRL-QAS}  

\begin{document}

\maketitle

\begin{abstract}
Reinforcement learning (RL) emerges as a prominent approach to quantum architecture search (QAS), the automated design of quantum circuits. However, transferring what an agent has learned from one number of qubits to another is still a weakly explored area. The state, the action space and the neural network of a state-of-the-art RL-QAS agent all grow with the number of qubits, so transfer learning across different qubit dimensions becomes prohibitive. In this article we make this transfer possible by introducing \textbf{\method{}}, a latent space transfer and refresh framework for RL-based QAS. In the core architecture of \method{} we utilize a circuit variational autoencoder, which encodes a circuit as one token per gate and maps circuits on any number of qubits into the same latent space. An RL agent then navigates this space instead of adding gates, so its state, its actions and its network have the same size at every qubit count. The latent space is periodically refreshed with the new agent's discoveries and, for a larger system, grown from the solutions of the smaller one. What is transferred is therefore the representation of circuits, not network weights or past experience. To show transfer, we consider learning among absolutely maximally entangled (AME) states, which are a resource for quantum error correction. \method{} discovers the five- and six-qubit AME states without being shown them and prepares them with $89$-$99\%$ of evaluation episodes, whereas the baseline RL-QAS fails to solve it with the same budget even with a network of size $10$ to $17$ times larger. We show the robustness of \method{} under depolarizing noise in which single-qubit gate errors are amplified $10\times$ and two-qubit gate errors $5\times$ compared with IBM quantum hardware. Hence by decoupling the RL agent from number of qubits, \method{} helps quantum architecture search to reuse quantum circuit representation learned on smaller problems. This provides a foundation to the transferable design of circuits across different quantum optimization problems.
\end{abstract}

\section{Introduction}
\label{sec:introduction}
Quantum computers promise to solve problems that are intractable for classical machines, from simulating molecules~\cite{aspuruguzik2005molecular,kassal2011chemistry,peruzzo2014vqe,omalley2016scalable,cao2019chemistry,mcardle2020chemistry} and materials~\cite{bauer2020materials} to sampling from distributions that classical computers cannot reproduce~\cite{bouland2019rcs,arute2019supremacy,madsen2022advantage}. One of the most prominent ways to do quantum computation is through quantum circuits, a sequence of gates applied to a register of qubits~\cite{nielsen2010quantum}, and many of the most promising near-term applications, such as variational eigensolvers and the preparation of entangled resource states, reduce to finding a circuit that prepares a particular state~\cite{cerezo2021vqa,peruzzo2014vqe}. On current noisy intermediate-scale quantum (NISQ) hardware this is difficult~\cite{preskill2018nisq}. Qubits decohere quickly, two-qubit gates are error-prone, and qubits are connected only to a few neighbours. So, a useful circuit must be short, use few entangling gates, and respect the device's connectivity. Designing such circuits by hand quickly becomes impractical as the number of qubits grows~\cite{plesch2011stateprep,cerveralierta2019ame}.

The common way to avoid designing each circuit by hand is to fix its structure in advance: one chooses a template, called an ansatz, which determines which gates act on which qubits, and optimises only the parameters of these gates. Hardware-efficient ansätze~\cite{kandala2017hardware} are friendly to the quantum hardware: starting from the native gate set and the qubit connectivity of a device, one builds a circuit of alternating layers of single-qubit rotations and entangling gates. They are shallow by design, but their structure is generic and not adapted to the problem, and whether shallow hardware-efficient ansätze can be trained and are useful depends strongly on the task: a favourable ``Goldilocks'' scenario has been identified only for inputs whose entanglement follows an area law~\cite{leone2024hea}. Problem-inspired ansätze, such as the unitary coupled-cluster ansatz with single and double excitations (UCCSD) of quantum chemistry~\cite{peruzzo2014vqe,romero2019ucc,anand2022ucc}, encode knowledge of the target, but their circuits are deep. The fermionic operators of the problem must be mapped to qubit operators, and the weight of these operators, the number of qubits each of them acts on, grows with the number of qubits $n$: as $O(n)$ in the Jordan-Wigner encoding~\cite{jordan1928paulische} and as $O(\log n)$ in the Bravyi-Kitaev encoding~\cite{bravyi2002fermionic,seeley2012bk}. Every such operator becomes a sequence of entangling gates of corresponding length, which makes the circuits that prepare the ground state deep and unfriendly to near-term hardware. In both cases the structure is chosen before anything is known about the solution, so the circuit is not adapted to the target it has to prepare. Hence, one way to progress in the search for ansätze that are hardware-aware and yet effective is to design the structure of the circuit automatically, with advanced classical search algorithms of the kind widely used in neural architecture search~\cite{zoph2017nas,elsken2019nas}. Inspired by this, \emph{quantum architecture search} (QAS) was introduced: it searches for the structure of the circuit itself~\cite{martyniuk2024qassurvey}, with adaptive~\cite{grimsley2019adapt}, differentiable~\cite{zhang2022dqas,wu2023quantumdarts,kumar2025rhodarts}, evolutionary~\cite{ding2022eqas}, Bayesian~\cite{nicoli2023bayesian}, tree-search~\cite{wang2023mcts} and predictor-based~\cite{zhang2021neuralpredictor,he2023gsqas} methods. These methods find useful circuits, but each search is typically run from scratch for each problem and each system size, and the cost of the search grows quickly with the number of qubits.

Reinforcement learning (RL) is a natural fit for QAS: an agent builds a circuit gate by gate and is rewarded for the quality of the result~\cite{ostaszewski2021rl,patel2024crlqas,kundu2024kanqas}. Quality has two sides: the accuracy of the circuit, such as its fidelity with a target state or its error in a ground-state energy, and its cost, such as its depth and its number of two-qubit gates. The two pull in opposite directions, since a deeper circuit can be more accurate but is harder to run. Reward functions can nevertheless be engineered to serve both goals at once: QASER~\cite{moflic2025qaser} takes these seemingly contradictory objectives into account in the reward and obtains circuits that are both shallower and more accurate, which shows that RL can handle rewards with competing aims. RL-based QAS has produced compact, noise-robust circuits for variational problems~\cite{patel2024crlqas}, state preparation~\cite{kuo2021qasdrl} and fault-tolerant protocols~\cite{zen2025faulttolerant}. It has three well-known drawbacks. First, the number of agent-environment interactions needed grows steeply with the number of qubits, because the agent's state encoding and action space grow with it. Second, performance depends strongly on the choice of state encoding, reward, agent and network size, even for simple problems~\cite{ikhtiarudin2025benchrl}. Third, an agent trained on one problem is of little use for another. TensorRL-QAS addresses scaling by warm-starting the agent with a tensor-network approximation of the target, reaching $20$ qubits~\cite{kundu2025tensorrl}, but this relies on domain knowledge about the target rather than on what was learned before. How to \emph{transfer} what an RL agent has learned from one problem to another remains an open question. Almost every RL-QAS agent is trained from scratch. Two forms of transfer have been tried. The first reuses a trained network~\cite{zen2025faulttolerant}. This requires the state and action spaces of the two problems to match, which they do not when the number of qubits changes. The second transfers the replay buffer~\cite{kundu2026replay}. This carries over experience, but not a representation that generalises across system sizes.

\begin{figure}[t!]
    \centering
    \caption{\small\textbf{Overview of \method{}}. (a) A circuit is encoded as one token per gate, as defined in Eq.~\eqref{eq:token}, and mapped by a bidirectional GRU to a latent point $z\in\mathbb{R}^8$, a fixed representation whose size does not depend on the number of qubits. (b) A DDQN agent moves the latent point through the latent space while the VAE is kept fixed. Each move is $z\to z\pm\delta e_k$: a step of size $\delta$ along one latent axis, where $e_k$ is the unit vector of axis $k$. At every point 
    the agent is rewarded according to 
    Eq.~\eqref{eq:log-reward}. (c) After each block of $M$ episodes, the solutions and best partial solutions found are added to a pool and the VAE is fine-tuned on it
    (d) Along the ladder of targets, each level's latent space is built from the previous level's solutions, which are relabelled, reordered and grown into a transfer pool. 
    The distribution (in blue) in the latent space does not represent the real data distribution; it is for illustrative purposes only.}
    \includegraphics[width=0.8\linewidth]{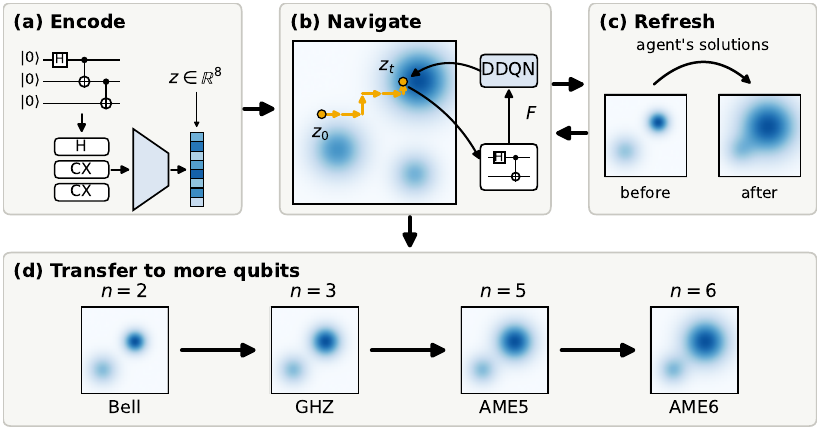}
    \label{fig:pipeline}
\end{figure}

In this work we make the \emph{representation} of circuits the object of transfer. Our approach is motivated by work in classical machine learning in which search and control take place in the latent space of a generative model rather than in the original space. Optimising in the latent space of a variational autoencoder (VAE)~\cite{kingma2014vae,kingma2019introduction} enabled gradient-based molecular design~\cite{gomezbombarelli2018automatic}, retraining the model on the points found makes this search markedly more effective~\cite{tripp2020weighted}, unsupervised latent embeddings of neural architectures make RL and Bayesian architecture search more sample efficient~\cite{yan2020arch2vec}, and offline RL agents act in the latent space of a VAE so that every decoded action is valid~\cite{zhou2020plas}. We bring these ideas to RL-based QAS and introduce our framework, \method{} (latent reinforcement learning for quantum architecture search), illustrated in Fig.~\ref{fig:pipeline}. The framework contains 4 unique sub-modules in which the input of each next step is dependent on the output of the previous. In the first step Fig.~\ref{fig:pipeline}(a) we encode circuits at the level of operations, one token per gate, and train a circuit VAE whose latent space has a fixed dimension $d=8$ for every number of qubits. A double deep Q-network (DDQN) agent~\cite{vanhasselt2016ddqn} in step Fig.~\ref{fig:pipeline}(b)then navigates this space in fixed steps: it observes only the eight coordinates of its current latent point and chooses among 16 moves, a step forward or back along each axis, whether the circuit acts on two qubits or six, and every point it visits is decoded into a circuit and simulated. Following the previous step in Fig.~\ref{fig:pipeline}(c) the latent space is periodically refreshed with the circuits the agent discovers, and the space for $n$ qubits is grown from the solutions found at a smaller qubit count, so that what is carried between system sizes is a representation of circuits, not a network whose shape depends on $n$.Finally in step Fig.~\ref{fig:pipeline}(d) we observe the transfer learning is taking place across different qubit sizes. The complete latent representation based RL-QAS transfer framework is elaborately discussed in Sec.~\ref{sec:method}. In summary we make three following contributions:
\begin{enumerate}
    \item \textbf{\method{}, a latent space transfer learning framework for RL-based QAS.} The agent acts in the latent space of a circuit VAE instead of on the circuit itself. The latent space is periodically refreshed with the circuits the agent discovers, and the latent space for a larger system is built from the solutions found for a smaller one, so that what is transferred between system sizes is the representation. The framework is noise aware: every fidelity it uses can be evaluated under gate noise, and with depolarising noise on every gate it finds and learns all targets at rates comparable to the noiseless case.
    \item \textbf{A circuit encoding with constant qubit overhead.} Within this framework we introduce an operation-level encoding of quantum circuits, one token per gate, whose size depends on the number of gates and not on the number of qubits. The VAE built on it grows by $2\%$ from two to ten qubits and the agent not at all, whereas the network of a gate-based agent grows thirty-fold. This is what makes transfer across qubit counts possible: the same latent dimension, and the same agent, serve every system size.
    \item \textbf{An analysis of the latent space and a comparison with gate-based baselines.} We evaluate the framework on a ladder of increasingly entangled state-preparation targets: the Bell states, the GHZ state, and the absolutely maximally entangled (AME) states on five and six qubits. The latent space uses all of its eight dimensions at every qubit count. \method{} finds the AME states on five and six qubits without the target ever being supplied, and learns to prepare them in $89$-$99\%$ and $96$-$99\%$ of evaluation episodes. A gate-based DDQN with the same budget solves neither target in any of $30$ runs, and with the same transferred circuits or its own curriculum it finds the targets at times but does not learn to prepare AME6. The refreshed latent space holds hundreds of distinct circuits for each target, so that together with the policy it acts as an on-demand generator of target-preparing circuits.
\end{enumerate}

The rest of the paper is organised as follows. Section~\ref{sec:preliminaries} introduces state preparation, QAS, variational autoencoders and DDQN, together with the related work on each. Section~\ref{sec:method} presents the method, Sec.~\ref{sec:setup} the experimental setup, and Sec.~\ref{sec:results} the results. We discuss the results in Sec.~\ref{sec:discussion} and conclude in Sec.~\ref{sec:conclusion}. Limitations and future work are the subject of Sec.~\ref{sec:limitations}.

\FloatBarrier  
\section{Preliminaries}
\label{sec:preliminaries}

This section fixes notation and summarises the three ingredients our method combines. Section~\ref{sec:prelim-quantum} introduces quantum state preparation with parameterised circuits, and Sec.~\ref{sec:prelim-qas} the search for good circuit structures. Section~\ref{sec:prelim-vae} covers variational autoencoders, the latent spaces they induce, and earlier work that searches in such spaces. Sections~\ref{sec:prelim-rl} and~\ref{sec:prelim-ddqn} cover value-based deep reinforcement learning, in particular Double DQN. How these pieces are put together is the subject of Sec.~\ref{sec:method}.

\subsection{Quantum circuits and state preparation}
\label{sec:prelim-quantum}

An $n$-qubit pure state is a unit vector $\ket{\psi}\in\mathcal{H}=(\mathbb{C}^2)^{\otimes n}$, and a quantum circuit is a finite sequence of gates $C=(g_1,\dots,g_T)$, each acting on one or two qubits, that implements the unitary $U_C=U_{g_T}\cdots U_{g_1}$~\cite{nielsen2010quantum}. Throughout this work gates are drawn from the discrete set
\begin{equation}
    \mathcal{G}=\{\,H,\;X,\;Z,\;\mathrm{CX},\;\mathrm{CZ}\,\},
    \label{eq:gate-set}
\end{equation}
where $\mathrm{CX}$ and $\mathrm{CZ}$ are the controlled-NOT and controlled-$Z$ gates. No gate carries a continuous parameter, so a gate is fully specified by its type and one or two qubit indices, and a circuit is a variable-length sequence of such discrete tokens. This is the structure the circuit encoder of Sec.~\ref{sec:method} has to represent: it maps every such sequence, whatever its length or number of qubits, to a vector of fixed size, so that the search for good circuits can take place in a continuous space whose dimension does not depend on $n$. The set $\mathcal{G}$ consists of Clifford gates but lacks the phase gate $S$, so it is neither universal nor able to generate the full Clifford group. It is nevertheless sufficient for every target considered here: each is a graph state
\begin{equation}
    \ket{G}=\prod_{(i,j)\in E}\mathrm{CZ}_{ij}\,H^{\otimes n}\ket{0}^{\otimes n},
    \label{eq:graph-state}
\end{equation}
defined by a graph $G=(V,E)$ on the $n$ qubits, up to $H$, $X$ or $Z$ gates on individual qubits~\cite{hein2004graph}, and can therefore be prepared exactly within $\mathcal{G}$. We focus on graph states because they are a central resource in quantum information: they are the starting point of measurement-based quantum computation~\cite{raussendorf2001oneway}, they are equivalent to the code states of stabiliser quantum error-correcting codes~\cite{hein2004graph}, and they include the most strongly entangled multi-qubit states, such as the AME states introduced below. Beyond this, Ref.~\cite{paraskevopoulos2025spin} recently prepared AME states on near-term spin-qubit architectures with exactly such circuits, applying CZ gates to qubits initialised in $\ket{+}$. This is why $\mathrm{CZ}$, the gate that creates the entanglement between neighbouring vertices of a graph state, is included in $\mathcal{G}$.

\paragraph{State preparation.}
Given a target state $\ket{\psi^\star}$, the state preparation task is to find a circuit $C$ whose output $\ket{\psi_C}=U_C\ket{0}^{\otimes n}$ is close to the target, as measured by the fidelity
\begin{equation}
    F(C)=\bigl|\braket{\psi^\star|\psi_C}\bigr|^2\in[0,1].
    \label{eq:fidelity}
\end{equation}
We call a circuit a \emph{solution}, or \emph{good} circuit, when $F(C)\ge\xi$ for an acceptance threshold $\xi$ close to one. When a family of equivalent targets is acceptable, for instance any of the four Bell states for $n=2$, which are defined in App.~\ref{app:bell}, $F$ is taken as the maximum over that family. On noisy intermediate-scale quantum (NISQ) hardware~\cite{preskill2018nisq}, a good circuit must also be cheap to run. It should have low depth, since qubits decohere within a limited time; it should use few two-qubit gates, whose error rates are typically about an order of magnitude higher than those of single-qubit gates; and its two-qubit gates should respect the hardware's native connectivity, since gates between non-adjacent qubits must be routed through additional SWAP gates. Searching for a circuit structure over $\mathcal{G}$ that meets these requirements is an instance of \emph{quantum architecture search} (QAS), which we review in Sec.~\ref{sec:prelim-qas}.

\paragraph{Target states.}
We use targets of increasing entanglement complexity. For $n=2$ these are the Bell states $\ket{\Phi^\pm}=(\ket{00}\pm\ket{11})/\sqrt2$ and $\ket{\Psi^\pm}=(\ket{01}\pm\ket{10})/\sqrt2$, and for larger $n$ the Greenberger-Horne-Zeilinger (GHZ) state $(\ket{0}^{\otimes n}+\ket{1}^{\otimes n})/\sqrt2$~\cite{greenberger1989ghz}. The most demanding targets are \emph{absolutely maximally entangled} (AME) states, for which every subset of $\lfloor n/2\rfloor$ qubits is maximally mixed, so that the state is maximally entangled across every bipartition~\cite{scott2004multipartite}. Qubit AME states exist only for $n\in\{2,3,5,6\}$: an $n$-qubit AME state is equivalent to a quantum error-correcting code that encodes no logical qubit but detects the largest possible number of errors, and such codes are impossible for $n=4$~\cite{higuchi2000couples}, $n=7$~\cite{huber2017seven} and $n\ge8$~\cite{scott2004multipartite}. For $n=5,6$ they can be written directly as graph states of the form~\eqref{eq:graph-state}~\cite{hein2004graph}. Because they are maximally entangled in this strong sense, AME states have been used as benchmarks of the multipartite entangling capability of near-term hardware~\cite{paraskevopoulos2025spin}; for the same reason we use them as a ladder of state preparation targets of increasing difficulty for $n=2,3,5,6$; the details are given in Sec.~\ref{sec:setup}.

\subsection{Quantum architecture search}
\label{sec:prelim-qas}

Variational quantum algorithms prepare a target state or minimise a cost function with a parameterised circuit whose structure, the \emph{ansatz}, is usually fixed in advance~\cite{cerezo2021vqa}. Two families of fixed ansätze dominate. Hardware-efficient ansätze alternate layers of single-qubit rotations and entangling gates, chosen from the native gate set of a device and placed according to its qubit connectivity~\cite{kandala2017hardware}; they are shallow and friendly to the hardware but problem-agnostic, and their usefulness depends strongly on the task~\cite{leone2024hea}. Problem-inspired ansätze, such as the UCCSD ansatz of quantum chemistry~\cite{peruzzo2014vqe,romero2019ucc,anand2022ucc}, build in knowledge of the target but yield circuits too deep for current hardware, because the weight of the qubit operators that represent the fermionic problem grows with the number of qubits, as $O(n)$ in the Jordan-Wigner and $O(\log n)$ in the Bravyi-Kitaev encoding~\cite{jordan1928paulische,bravyi2002fermionic,seeley2012bk}. A fixed structure is therefore either too shallow to express the target, or deeper and noisier than necessary. \emph{Quantum architecture search} (QAS) instead treats the structure itself as the object of optimisation: which gates are applied, on which qubits, and in which order~\cite{martyniuk2024qassurvey,kundu2024rlqasthesis}.

Many search strategies have been proposed. Adaptive methods grow the circuit one operator at a time from a predefined pool, guided by energy gradients~\cite{grimsley2019adapt,tang2021qubitadapt}. Differentiable methods relax the discrete choice of gates into a continuous distribution that can be optimised by gradient descent~\cite{zhang2022dqas,wu2023quantumdarts,kumar2025rhodarts}. Other approaches use evolutionary algorithms~\cite{ding2022eqas}, Bayesian optimisation~\cite{nicoli2023bayesian}, Monte Carlo tree search~\cite{wang2023mcts}, weight sharing in a supernet~\cite{du2022quantumsupernet}, neural predictors of circuit performance~\cite{zhang2021neuralpredictor}, noise-adaptive search~\cite{wang2022quantumnas}, and training-free proxies~\cite{he2024trainingfree}. These methods show that useful circuit structures can be found automatically, but scaling to larger systems and robustness to hardware noise remain open problems~\cite{martyniuk2024qassurvey,kundu2025tensorrl}.

\paragraph{Reinforcement learning for QAS.}
Reinforcement learning, introduced in Sec.~\ref{sec:prelim-rl}, is a natural fit for QAS, because a circuit can be built gate by gate as a sequence of decisions, with the quality of the result as the reward. In the standard formulation~\cite{ostaszewski2021rl,patel2024crlqas,kundu2024kanqas}, each episode starts from an empty circuit. The state is an encoding of the circuit built so far, typically a binary tensor of size $L\times(n+k)\times n$: for each of $L$ circuit layers, an $n\times n$ block records the two-qubit gates by their control and target qubits, and a $k\times n$ block records which qubits carry one of $k$ single-qubit gate types, the row giving the type or, for rotations, the axis~\cite{patel2024crlqas}. Each action appends one gate from a discrete set: a two-qubit gate given by its control and target qubits, or a single-qubit gate given by its type and qubit, so there are $O(n^2)$ actions. The reward is computed from the cost of the current circuit, here the fidelity of Eq.~\eqref{eq:fidelity}, usually with a large bonus once the cost passes the acceptance threshold; the episode ends at that point or after a budget of $L$ gates. Common refinements mask actions that would immediately undo the previous gate, tighten the acceptance threshold through a curriculum as the agent improves~\cite{ostaszewski2021rl,patel2024crlqas}, and prune redundant gates~\cite{patel2024crlqas}.

Within this framework, RL agents have been used to design ansätze for variational ground-state problems~\cite{ostaszewski2021rl,patel2024crlqas}, to prepare multi-qubit entangled states~\cite{kuo2021qasdrl}, to optimise and compile circuits~\cite{fosel2021circuitopt,moro2021compiling,kremer2024transpiling}, and to discover fault-tolerant state preparation circuits~\cite{zen2025faulttolerant}. Recent work has improved such agents along several axes: curriculum learning and pruning for robustness to hardware noise (CRLQAS)~\cite{patel2024crlqas}, Kolmogorov-Arnold networks as function approximators (KANQAS)~\cite{kundu2024kanqas}, learned composite gates~\cite{kundu2026gadgets}, tensor-network warm starts that reach 20 qubits (TensorRL-QAS)~\cite{kundu2025tensorrl}, and reliability-aware experience replay~\cite{kundu2026replay}. A benchmark of nine RL agents across QAS tasks finds that no single agent is best on every task~\cite{ikhtiarudin2025benchrl}.

Two limitations of RL-based QAS motivate this work. First, the state encoding, of size $O(Ln^2)$, and the action space, of size $O(n^2)$, both grow with the number of qubits, and the gate budget $L$ must grow with the complexity of the target. An agent trained at one qubit count therefore cannot be reused at another, the number of agent-environment interactions needed rises steeply with system size, and most RL-based QAS remains restricted to a few qubits~\cite{kundu2025tensorrl}. Second, almost every RL-based QAS agent is trained from scratch for each new problem, discarding what it learned before~\cite{kundu2026replay}. The existing forms of transfer move either the agent or its experience. Zen~\emph{et al.}~\cite{zen2025faulttolerant} reuse a trained agent's network on related state-preparation tasks, which requires the state and action spaces of the two tasks to match. Kundu and Feld~\cite{kundu2026replay} instead transfer the replay buffer, warm-starting noisy training with experience collected in noiseless training, without transferring network weights. The tensor-network warm start of TensorRL-QAS~\cite{kundu2025tensorrl} is a form of domain knowledge about the target rather than transfer between problems. In Sec.~\ref{sec:method} we instead let the agent act in a VAE latent space of fixed dimension $d$, and transfer this representation rather than the agent: the latent space for a larger system is built from the circuits discovered for a smaller one, while the agent itself, whose input and output sizes depend only on $d$, can be carried over or retrained.

\subsection{Variational autoencoders and latent spaces}
\label{sec:prelim-vae}

A variational autoencoder (VAE)~\cite{kingma2014vae,rezende2014stochastic,kingma2019introduction} is a latent-variable generative model that learns a continuous, low-dimensional representation $z\in\mathbb{R}^d$ of structured data $x$. It consists of a probabilistic \emph{encoder} $q_\phi(z\mid x)=\mathcal{N}\bigl(z;\mu_\phi(x),\operatorname{diag}\sigma^2_\phi(x)\bigr)$, a \emph{decoder} $p_\theta(x\mid z)$, and a prior $p(z)=\mathcal{N}(0,I_d)$. The encoder and decoder are trained jointly by minimising the (negative) evidence lower bound, which we use in its $\beta$-weighted form~\cite{higgins2017betavae},
\begin{equation}
    \mathcal{L}_\beta(\phi,\theta;x)=
    \underbrace{\mathbb{E}_{q_\phi(z\mid x)}\bigl[-\log p_\theta(x\mid z)\bigr]}_{\text{reconstruction}}
    +\beta\,\underbrace{D_{\mathrm{KL}}\bigl(q_\phi(z\mid x)\,\|\,p(z)\bigr)}_{\text{regularisation}} .
    \label{eq:elbo}
\end{equation}
The expectation is estimated with reparameterisation $z=\mu_\phi(x)+\sigma_\phi(x)\odot\varepsilon$, $\varepsilon\sim\mathcal{N}(0,I_d)$, which makes the objective differentiable in $\phi$~\cite{kingma2014vae}. For a Gaussian posterior and prior the KL term has the closed form $\tfrac12\sum_{k=1}^d\bigl(\mu_k^2+\sigma_k^2-\log\sigma_k^2-1\bigr)$.
\begin{figure}[h!]
    \centering
    \caption{\small\textbf{Searching for circuits in a VAE latent space (schematic)}. A circuit $x$ with low target fidelity is encoded to a latent point $z_0$. A sequence of steps moves the point through the latent space, whose shading indicates the fidelity of the circuit each point decodes to (darker is higher), into a high-fidelity region at $z_T$. Decoding $z_T$ yields a circuit that prepares the target state, here the Bell state $\ket{\Phi^+}$. The landscape is illustrative, not data.}
    \includegraphics[width=0.8\linewidth]{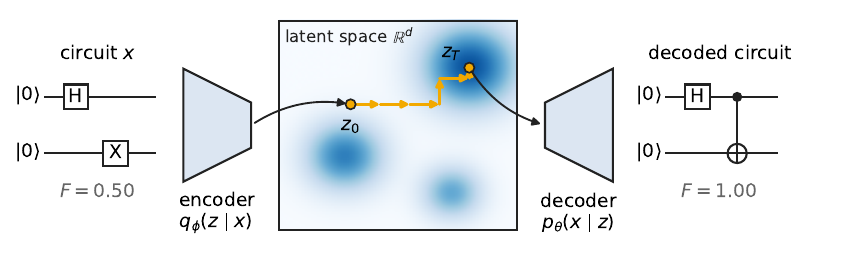}
    \label{fig:latent-schematic}
\end{figure}

The two terms of Eq.~\eqref{eq:elbo} pull in opposite directions. The \emph{reconstruction} term rewards the encoder for storing as much information about $x$ in $z$ as possible, so that the decoder can recover $x$ exactly. The \emph{regularisation} term rewards the encoder for placing every input close to the origin, in the same standard Gaussian cloud. On its own, the first term would scatter the inputs to isolated, far-apart points; on its own, the second would collapse them all onto the same point. The weight $\beta$ sets the compromise between them, and at a suitable value the result is a latent space with three useful properties: similar inputs are placed close together, the space around the origin is filled without gaps, and a small step from any point lands on a point that still decodes to a sensible input. Such a space can be \emph{navigated}: one can move through it in small steps and, at every point, decode the corresponding object. This idea underlies latent space optimisation in molecular design~\cite{gomezbombarelli2018automatic} and is the property our agent exploits.

\paragraph{Sequence VAEs.}
When $x$ is a variable-length sequence of discrete tokens $(x_1,\dots,x_T)$, encoder and decoder are typically recurrent networks~\cite{bowman2016generating}. A (bidirectional) recurrent encoder, here a gated recurrent unit (GRU)~\cite{cho2014gru}, summarises the sequence into $(\mu_\phi,\log\sigma^2_\phi)$, and an autoregressive decoder factorises
\begin{equation}
    p_\theta(x\mid z)=\prod_{t=1}^{T}p_\theta\bigl(x_t\mid x_{<t},z\bigr),
    \label{eq:autoregressive}
\end{equation}
so that the reconstruction term becomes the cross entropy summed over all tokens of the sequence. At inference time a sequence is produced from a latent point by greedy decoding, $\hat x_t=\arg\max_{x_t}p_\theta(x_t\mid\hat x_{<t},z)$, which turns the decoder into a deterministic map $\mathrm{Dec}:\mathbb{R}^d\to\{\text{circuits}\}$. Autoregressive decoders are known to ignore $z$ if the KL term dominates early in training; a standard remedy, which we adopt, is to anneal $\beta$ from zero to its final value~\cite{bowman2016generating}.

Figure~\ref{fig:latent-schematic} summarises how such a latent space is used for circuit search. An encoder maps a circuit to a latent point $z_0$; a search procedure moves through the latent space in steps; and the decoder maps the final point $z_T$ back to a circuit. If good circuits occupy connected regions of the latent space, a short sequence of steps that starts from a poor circuit can end inside such a region, and decoding the end point then yields a good circuit. In our method these steps are taken by an RL agent, as discussed in detail in Sec.~\ref{sec:method}.

\paragraph{Learned representations for architecture search.} In classical neural architecture search, several methods first map discrete architectures into a continuous space and then search in that space. Neural Architecture Optimisation trains an encoder, a decoder and a performance predictor jointly, and improves architectures by gradient steps on their continuous embeddings~\cite{luo2018nao}. D-VAE~\cite{zhang2019dvae} and arch2vec~\cite{yan2020arch2vec} learn the space with a graph variational autoencoder; arch2vec in particular trains it without any performance labels and shows that the resulting smooth latent space makes reinforcement learning and Bayesian optimisation search markedly more efficient than search over discrete encodings. These ideas have recently been brought to QAS. GSQAS pre-trains a graph encoder of quantum circuits on unlabeled circuits to reduce the number of labelled circuits a performance predictor needs~\cite{he2023gsqas}. Closest to our work, Sun~\emph{et al.}~\cite{sun2026qasurl} encode quantum circuits as directed acyclic graphs, train a graph VAE on randomly generated circuits without labels, and then search the frozen latent space with REINFORCE and Bayesian optimisation, finding high-fidelity state preparation and VQE circuits for up to 12 qubits in fewer evaluations than predictor-based methods. We build on this idea but differ in three respects. First, our agent does not select latent points in one shot but navigates the latent space in fixed steps, with the circuit decoded at every step. Second, the latent space is not fixed after pre-training: it is refreshed with the circuits the agent discovers. Third, the latent dimension does not depend on the number of qubits, which lets us transfer the latent space between system sizes, whereas Ref.~\cite{sun2026qasurl} trains a separate model for each qubit count.

\paragraph{Retraining and navigating a latent space.} Two further ideas from classical machine learning enter our method. The first concerns the content of the latent space. Searching the latent space of a generative model instead of the discrete object space was popularised in molecular design, where a VAE over molecular strings enables gradient-based and Bayesian optimisation of chemical properties~\cite{gomezbombarelli2018automatic}. A known limitation is that a latent space trained once, on generic data, represents high-scoring objects poorly, because they are rare in its training set. Weighted retraining addresses this by periodically retraining the generative model on the points found during the search, weighted by their score~\cite{tripp2020weighted}. Our periodic refresh follows the same principle, retraining the VAE on the agent's discoveries with an increased weight, here inside an RL loop. The second concerns who moves through the latent space. Letting an RL agent act in the latent space of a generative model, rather than in the original action space, has been used in classical control: PLAS, for example, trains an offline RL policy in the latent space of a conditional VAE, so that every decoded action stays within the support of the data~\cite{zhou2020plas}. We use the same mechanism for a different purpose: the decoder guarantees that every latent point corresponds to a valid circuit, and the fixed latent dimension decouples the size of the agent's state and action spaces from the number of qubits.

\FloatBarrier  
\subsection{Reinforcement learning}
\label{sec:prelim-rl}

Reinforcement learning (RL) studies an agent that learns to act by interacting with an environment~\cite{sutton2018rl}. The interaction is modelled as a Markov decision process (MDP) $(\mathcal{S},\mathcal{A},P,r,\gamma)$ with state space $\mathcal{S}$, action space $\mathcal{A}$, transition kernel $P(s'\mid s,a)$, reward function $r(s,a,s')$ and discount factor $\gamma\in[0,1)$. At each step $t$ the agent observes $s_t$, chooses $a_t\sim\pi(\cdot\mid s_t)$ according to its policy $\pi$, and receives $r_{t+1}$ as the environment moves to $s_{t+1}$. In the episodic setting used here an episode ends after a fixed budget of $L$ steps or earlier when a terminal condition is met. The agent's goal is to maximise the expected return $G_t=\sum_{k=0}^{L-t-1}\gamma^k r_{t+k+1}$.

The action-value function $Q^\pi(s,a)=\mathbb{E}_\pi[G_t\mid s_t=s,a_t=a]$ measures how good it is to take action $a$ in state $s$ and follow $\pi$ thereafter. The optimal action-value function $Q^\ast=\max_\pi Q^\pi$ satisfies the Bellman optimality equation
\begin{equation}
    Q^\ast(s,a)=\mathbb{E}_{s'\sim P(\cdot\mid s,a)}\Bigl[r(s,a,s')+\gamma\max_{a'\in\mathcal{A}}Q^\ast(s',a')\Bigr],
    \label{eq:bellman}
\end{equation}
and any policy that is greedy with respect to $Q^\ast$ is optimal. Value-based methods therefore learn an estimate $Q\approx Q^\ast$ and act (approximately) greedily on it. To keep exploring during learning, actions are chosen $\epsilon$-greedily: a uniformly random action with probability $\epsilon$ and $\arg\max_aQ(s,a)$ otherwise, with $\epsilon$ decayed from $1$ towards a floor $\epsilon_{\min}$ as training proceeds.

For the RL-algorithm we utilise a Double Deep Q-network (DDQN) agent which contains a fully connected network $Q_\vartheta$ with the ADAM optimiser~\cite{kingma2015adam} on Eq.~\eqref{eq:dqn-loss}, using the target of Eq.~\eqref{eq:ddqn-target}; the target network is a hard copy of the online network, refreshed every $K$ steps. The discount factor is tied to the episode budget as $\gamma=\gamma_{\mathrm{f}}^{1/L}$, so that the reward at the final step of an episode is discounted by $\gamma_{\mathrm{f}}$ regardless of $L$. The whole DDQN learning algorithm is discussed in App.~\ref{sec:prelim-ddqn}. As for the network we use a multilayer perceptron (MLP) with ReLU activations. All hyperparameter and their values are provided in App.~\ref{app:hyperparameters}.

DDQN is the agent used in several RL-QAS methods, including CRLQAS~\cite{patel2024crlqas}, KANQAS~\cite{kundu2024kanqas} and TensorRL-QAS~\cite{kundu2025tensorrl}, which makes our results directly comparable to that line of work. The BenchRL-QAS benchmark~\cite{ikhtiarudin2025benchrl}, which compares nine RL agents on QAS tasks of 2 to 8 qubits, finds that no single agent is best on every task and metric. Within that picture, DDQN is among the agents that most consistently produce shallow circuits with few gates, and it is competitive in run time. We therefore use DDQN for every experiment, so that differences between methods come from the representation (latent space versus gate-by-gate actions) and not from the choice of learner.

\section{Efficient transfer via latent space representation}
\label{sec:method}

Transfer learning, reusing what was learned on one task to learn another faster, is well established in classical machine learning and reinforcement learning~\cite{taylor2009transfer,zhu2023transfer}. Its success depends on \emph{what} is transferred and \emph{how}~\cite{wei2018l2t}: network weights, stored experience, or a representation shared by the source and the target task. 
In RL-based QAS, transfer has so far moved either the agent's network~\cite{zen2025faulttolerant} or its experience~\cite{kundu2026replay}. Both generalise to an environment that resembles the original one, such as the noisy version of a problem first solved without noise~\cite{kundu2026replay}, but neither is robust when the new problem differs substantially from the old one.
This is because 
the RL-state described by a tensor-based encoding~\cite{patel2024crlqas, kundu2024vqsd} of the circuit that depends on the qubits the gates acts on. Hence both the agent's observable and the action space grows rapidly with the number of qubits. So a network or a replay buffer (under an off-policy algorithm) obtained at $n$ qubits therefore has no direct counterpart at $n'\neq n$: the transfer mechanism lacks robustness exactly when the qubit number differs, which is the direction in which problems become intractable.

In this section we introduce \method{} which addresses the bottleneck discussed above.
Our framework is suitable for transfer learning across qubit counts, in which what is learned on a small system carries over to a larger one. This requires a description of circuits whose size does not depend on the number of qubits. We achieve this by introducing a constant overhead encoding of quantum circuit based on latent space representation. 
In Sec.~\ref{sec:method-encoding} we provide the encoding of the circuit and show that the latent space naturally decouples the agent from the size of the circuit. Following this a gate-based agent observes and edits the circuit description itself, which grows with the number of qubits. Here, a circuit variational autoencoder, elaborated in Sec.~\ref{sec:method-vae}, maps every circuit
to a point in a latent space of fixed dimension $d$, and the agent observes and moves only in this space, so its observations and actions have the same size at every qubit count. A DDQN agent searches the space in fixed steps, as elaborated in Sec.~\ref{sec:method-mdp}, and the decoder then turns every point the agent visits back into a circuit through a refresh and transfer method described in Secs.~\ref{sec:method-refresh} and~\ref{sec:method-transfer}. 


\subsection{Constant qubit overhead circuit encoding}
\label{sec:method-encoding}

Transfer across qubit counts requires an encoding of circuits whose form does not change with $n$, so the central design choice is how a circuit is presented to the model. The standard RL-QAS encoding introduced in Sec.~\ref{sec:prelim-qas} is a binary tensor of size $L\times(n+k)\times n$. Each of its $L$ layers is a binary array with $n+k$ rows and $n$ columns: the first $n\times n$ block marks the two-qubit gates of the layer by their control and target qubits, and the remaining $k\times n$ block marks the single-qubit gates, where the row selects one of the $k$ gate types (for rotations, the axis of rotation) and the column the qubit the gate acts on. Its size is fixed by the number of qubits and the depth budget, most of its entries are zero, and a model built on it has to be rebuilt whenever $n$ changes.

We instead encode a circuit at the level of \emph{operations}. A circuit $C=(g_1,\dots,g_T)$ is the sequence of its gates, and each gate $g_t$ becomes one token
\begin{equation}
    \tau_t=\bigl(\,\mathrm{type}_t,\;q^{(1)}_t,\;q^{(2)}_t,\;t,\;\theta_t\,\bigr),
    \label{eq:token}
\end{equation}
with five fields. The first, $\mathrm{type}_t$, is the kind of gate, one element of the gate set $\mathcal{G}$ such as $H$ or $\mathrm{CX}$, stored as its index in $\mathcal{G}$. The next two, $q^{(1)}_t,q^{(2)}_t\in\{1,\dots,n\}$, are the qubits the gate acts on, the control and the target for a two-qubit gate; for a single-qubit gate, $q^{(2)}_t=0$ marks that there is no second qubit. The fourth, $t$, is the position of the gate in the sequence. The last, $\theta_t$, is the rotation angle of a parameterised gate such as $R_z(\theta)$, rescaled to $[-1,1]$. It is $0$ for a gate without a parameter, which is the case for every gate in $\mathcal{G}$, so this field is unused here.

\begin{figure}[h!]
    \centering
    \caption{\small \textbf{The scaling analysis of different kinds of quantum circuit encoding for RL-QAS}. (a) Both encodings, step by step, for the three-qubit GHZ circuit; the gate added in each step is highlighted. In the tensor encoding, on the left, every layer of the circuit has its own binary grid of $n+k$ rows and $n$ columns (here $n=3$, $k=5$), and a gate sets a single entry to $1$: in the upper block, the target (row) and control (column) of a two-qubit gate; in the lower block, the type (row) and qubit (column) of a single-qubit gate. In our encoding, on the right, every gate becomes one token with the five fields of Eq.~\eqref{eq:token}, whatever the number of qubits. (b) Size of the state as a function of the number of qubits, for the targets and gate budgets of Table~\ref{tab:sizes}, on a logarithmic scale. Solid orange: tensor encoding of gate-based RL-QAS, $L(n+k)n$ binary entries with $k=5$. Dotted grey: the same circuits as tokens, $5\,T_{\max}$ numbers. Dashed blue: the latent point observed by our agent, $d=8$. Filled markers are the levels studied in this work. Open markers ($n\ge8$) are computed from the same formulas for the targets planned as future work; no experiments were run there.}
    \includegraphics[width=\linewidth]{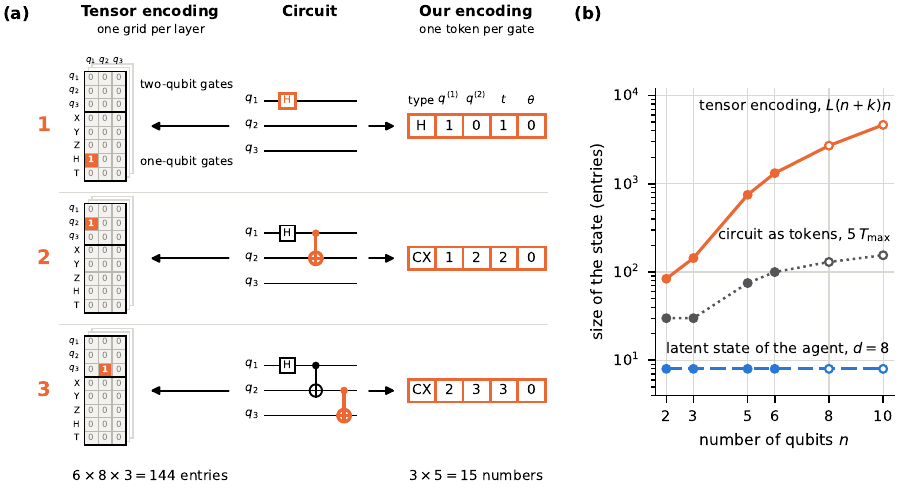}
    \label{fig:encoding-steps}
\end{figure}

For example, the three-qubit GHZ circuit that applies $H$ to qubit $1$, then $\mathrm{CX}$ from qubit $1$ to qubit $2$ and $\mathrm{CX}$ from qubit $2$ to qubit $3$ becomes the three tokens
\begin{equation*}
    \tau_1=(H,1,0,1,0),\qquad \tau_2=(\mathrm{CX},1,2,2,0),\qquad \tau_3=(\mathrm{CX},2,3,3,0),
\end{equation*}
$15$ numbers, against the $6\times8\times3=144$ binary entries the tensor encoding uses for the same circuit. Figure~\ref{fig:encoding-steps}(a) shows both encodings of this circuit, one gate at a time.

The five fields are embedded separately and concatenated,
\begin{equation}
    e_t=W_{\mathrm{out}}\bigl[\,E_{\mathrm{gate}}(\mathrm{type}_t)\,\|\,E_{q}^{(1)}(q^{(1)}_t)\,\|\,E_{q}^{(2)}(q^{(2)}_t)\,\|\,E_{\mathrm{pos}}(t)\,\|\,w_\theta\theta_t\,\bigr]\in\mathbb{R}^{64},
    \label{eq:token-embedding}
\end{equation}
with $16$-dimensional embedding tables $E$. A circuit is therefore a sequence of $T$ vectors of fixed width, and its length is the number of gates it contains, not the product of qubits and layers.

Figure~\ref{fig:encoding-steps}(b) compares the size of the two encodings for the targets of this paper. The tensor encoding grows from $84$ entries at $n=2$ to $1320$ at $n=6$ and $4650$ at $n=10$; the token sequence grows only with the number of gates; and the state the agent observes, the latent point of Sec.~\ref{sec:method-mdp}, has $8$ entries throughout.

This factorised encoding has three consequences that the rest of the method relies on.
\begin{enumerate}
    \item \textbf{Low overhead on the number of qubits.} The gate vocabulary, the recurrent networks and the latent space are shared by all qubit counts. Only the two qubit-index tables $E^{(1)}_q,E^{(2)}_q$ and the matching output heads of the decoder have $n+1$ entries, so the number of parameters grows by $O(n)$ with a small constant. Table~\ref{tab:sizes} shows the effect: from $n=2$ to $n=10$ the VAE grows by $2\%$, whereas the network of a gate-based agent grows thirty-fold.
    \item \textbf{The input to the agent is independent of $n$.} Because every circuit is summarised by a point $z\in\mathbb{R}^d$, the agent's observation has $d$ entries and its action set $2d$ elements for every number of qubits, as detailed in Sec.~\ref{sec:method-mdp}. With $d=8$, the same agent architecture, of $2626$ parameters, is used from Bell pairs to six-qubit AME states.
    \item \textbf{The size of the encoding scales with the number of gates, not of qubits.} A circuit of $T$ gates is described by $T$ tokens, independently of how many qubits are idle, so the encoder spends no capacity on the empty positions that dominate the tensor encoding at large $n$.
\end{enumerate}

\begin{table}[h!]
    \centering
    \small
    \caption{\textbf{Sizes of the representation and of the agents as a function of the number of qubits $n$}. The VAE uses the gate budget of Sec.~\ref{sec:setup} as its maximum sequence length; the gate-based agent uses the same budget as its number of layers $L$ and an MLP with three hidden layers of $30$ neurons, the layer sizes of KANQAS~\cite{kundu2024kanqas}, as does the latent agent. Entries for $n\ge7$ follow from the architectures; experiments there are future work.}
    \label{tab:sizes}
    \begin{tabular}{rrrrrrr}
        \toprule
        & \multicolumn{1}{c}{VAE} & \multicolumn{2}{c}{\textbf{Latent agent}} & \multicolumn{3}{c}{Gate-based agent} \\
        \cmidrule(lr){2-2}\cmidrule(lr){3-4}\cmidrule(lr){5-7}
        $n$ & parameters & obs./actions & parameters & observation & actions & parameters \\
        \midrule
        2  & 85\,842 & \textbf{8 / 16} & \textbf{2\,626} & 84    & 12  & 4\,782 \\
        3  & 86\,004 & \textbf{8 / 16} & \textbf{2\,626} & 144   & 21  & 6\,861 \\
        5  & 86\,472 & \textbf{8 / 16} & \textbf{2\,626} & 750   & 45  & 25\,785 \\
        6  & 86\,714 & \textbf{8 / 16} & \textbf{2\,626} & 1\,320 & 60  & 43\,350 \\
        8  & 87\,134 & \textbf{8 / 16} & \textbf{2\,626} & 2\,704 & 96  & 85\,986 \\
        10 & 87\,538 & \textbf{8 / 16} & \textbf{2\,626} & 4\,650 & 140 & 145\,730 \\
        \bottomrule
    \end{tabular}
\end{table}

\paragraph{Canonical gate order.}
Gates on disjoint qubits, and diagonal gates such as $\mathrm{CZ}$ and $Z$, commute, so many different sequences prepare the same state. For a sequence model such commutation rules act as noise: which of several equivalent gates comes next is arbitrary, and this makes the sequence difficult to predict. That the order in which data is presented strongly affects what a sequence model can learn, even when the order carries no meaning, is well documented~\cite{vinyals2016order}. Before a circuit is used for training we therefore rewrite it in a canonical order. Among the gates that can be applied next without changing the prepared state, we always take the one that comes first in a fixed ranking, by gate type ($H$, $X$, $Z$, $\mathrm{CX}$, $\mathrm{CZ}$) and then by qubit index. For graph-state circuits this places all Hadamard gates first, followed by the $\mathrm{CZ}$ gates sorted by the qubits they act on. The prepared state is unchanged. Without this step, a VAE trained on six-qubit graph-state circuits reconstructed fewer than $1\%$ of its training circuits; with it, $87\%$. The full analysis is given in App.~\ref{app:gate-order}.

\FloatBarrier  
\subsection{The circuit VAE}
\label{sec:method-vae}

The encoding of Sec.~\ref{sec:method-encoding} turns a circuit into the sequence of token vectors $e_1,\dots,e_T$ of Eq.~\eqref{eq:token-embedding}, one per gate, so its length differs from circuit to circuit. The agent, in contrast, needs a state of fixed size. The component introduced in this section closes this gap. It compresses the token sequence of any circuit into a single point $z\in\mathbb{R}^d$ and, in the other direction, turns any point $z$ back into a sequence of tokens, that is, into a circuit. Its input and its output are the token fields of Eq.~\eqref{eq:token}, and every circuit it is trained on is first written in the canonical gate order.

\paragraph{Why a VAE?}
Three requirements single out the variational autoencoder, introduced in Sec.~\ref{sec:prelim-vae}, for this role. First, the map must work in both directions. The agent moves through the latent space, and every point it visits has to be decoded into a circuit before it can be simulated and rewarded; an encoder alone, such as the learned embeddings used to predict circuit performance~\cite{he2023gsqas}, would not be enough. Second, the latent space must be smooth and free of holes. The agent moves in fixed steps from point to neighbouring point, so nearby points should decode to similar circuits, and the points between the encodings of the training circuits should decode to meaningful circuits as well. A plain autoencoder places no constraint on the space between its training points; the KL term of the VAE, which pulls all encodings towards a common prior, encourages exactly this structure~\cite{kingma2014vae,bowman2016generating}. Third, the prior gives the latent space a known distribution, from which we can draw the starting points of episodes and estimate which fraction of the space decodes to solutions. The same properties have made VAEs a standard choice for searching the latent spaces of molecules~\cite{gomezbombarelli2018automatic} and of neural architectures~\cite{zhang2019dvae}. Finally, a VAE can be fine-tuned on new circuits, which is what allows the latent space to follow the agent's discoveries, as described in Sec.~\ref{sec:method-refresh}.

\paragraph{Architecture.}
The circuit VAE is a sequence VAE over the tokens of Eq.~\eqref{eq:token}. The encoder is a bidirectional GRU with hidden size $64$ that reads the token vectors $e_1,\dots,e_T$ and whose final forward and backward states are mapped to $\mu_\phi(C)$ and $\log\sigma^2_\phi(C)$ in $\mathbb{R}^d$, $d=8$. The decoder is a GRU that receives $z$ both as its initial hidden state and, concatenated with the embedding of the previous token, at every step; four output heads predict the gate type, the two qubit indices and the angle of the next token, and an end-of-sequence token terminates the circuit. Decoding is greedy, so $\mathrm{Dec}:\mathbb{R}^d\to\{\text{circuits}\}$ is deterministic, and the agent always sees the same circuit at the same latent point. The decoded circuit is truncated at a maximum length $T_{\max}$, the gate budget of each target, which is specified in Sec.~\ref{sec:setup}.

\paragraph{Training.}
The VAE is trained on Eq.~\eqref{eq:elbo} with $\beta$ annealed linearly from $0$ to $0.25$ over the first $40\%$ of training~\cite{bowman2016generating}. Two further measures were necessary for the larger targets. An autoregressive decoder that is strong enough to model the training circuits on its own learns to ignore $z$: the posterior collapses onto the prior, all circuits are encoded to nearly the same point, and the latent point carries almost no information about the circuit. We observed exactly this at $n=6$, where an unregularised VAE reached a KL divergence of $0.002$ nats, reconstructed none of its training circuits, and decoded $1000$ random latent points to only $132$ distinct circuits; these results are reported in App.~\ref{app:collapse}.

We prevent the collapse with \emph{free bits}~\cite{kingma2016iaf}, which replace the KL term of each latent dimension by $\max\bigl(D_{\mathrm{KL},k},\lambda\bigr)$ with $\lambda=1$ nat, so that the encoder is never penalised for using a dimension up to that amount. Free bits make sure that the encoder writes information into $z$, but they do not force the decoder to read it. During training the decoder is given the true previous gates at every step, and in circuits with a regular structure these are often enough to guess the next gate without consulting $z$. When the agent uses the decoder, however, there are no true previous gates: the whole circuit has to be produced from the latent point alone. \emph{Word dropout}~\cite{bowman2016generating} closes this gap between training and use. It hides each input token of the decoder with probability $0.5$, which removes the information the decoder could otherwise lean on, so that the gates it cannot see must be taken from $z$. The effect on how the latent dimensions are used is analysed in Sec.~\ref{sec:results-representation}.

\subsection{Reinforcement learning in the latent space}
\label{sec:method-mdp}

The circuit VAE of Sec.~\ref{sec:method-vae} provides everything the agent's environment is built from. Its latent space $\mathbb{R}^d$ is the space the agent moves in, so a latent point takes the place of the gate-level description of a circuit as the agent's state. Its decoder $\mathrm{Dec}$ turns every point the agent visits into a circuit, which is then simulated to obtain the reward. Its prior $\mathcal{N}(0,I_d)$ and its encoder mean $\mu_\phi$ supply the points at which episodes start. The agent thus never handles gates or qubits directly: it searches for a circuit by searching for the latent point that decodes to it. The VAE is kept fixed while the agent acts, so that the same point always decodes to the same circuit. With these components, the agent solves a state-preparation task, as defined in Sec.~\ref{sec:prelim-quantum}, through the following decision process.

\paragraph{State.} The state is the current latent point $s_t=z_t\in\mathbb{R}^d$, an $8$-dimensional vector for every number of qubits.

\paragraph{Actions.} An action moves the point by a fixed step $\delta=0.5$ along one latent axis, $z_{t+1}=z_t\pm\delta\,e_k$ for $k=1,\dots,d$, giving $2d=16$ actions. The step size balances two effects. A step that is too small often decodes to the same circuit, so the move gives the agent no feedback; a step that is too large replaces most of the circuit, so the move is no longer local. With $\delta=0.5$, $88$-$94\%$ of the moves at $n=5$ and $n=6$ change the decoded circuit, and those that do change about a third to a half of its gates, while an episode can still cross $3$ to $5$ standard deviations of the prior along one axis. The measurement is reported in App.~\ref{app:step-size}; we did not tune $\delta$ further. Every action is always available and every latent point decodes to a valid circuit, so no action masking is needed, unlike the gate-based setting.

\begin{figure}[h!]
    \centering
    \caption{\small\textbf{Per-step reward as a function of the fidelity $F$ of the prepared state, for the fidelity reward $r=F$ and the graded reward of Eq.~\eqref{eq:log-reward}}. Solid line with circles: fidelity reward; dashed line with squares: graded reward. The markers show the only fidelities a stabiliser circuit can reach, $F=2^{-(n-m)}$ with $m$ stabilisers shared with the target, at which the graded reward equals $m/n$. Above the threshold $\xi=0.98$ (dotted) both rewards pay $50F$.}
    \includegraphics[width=0.8\linewidth]{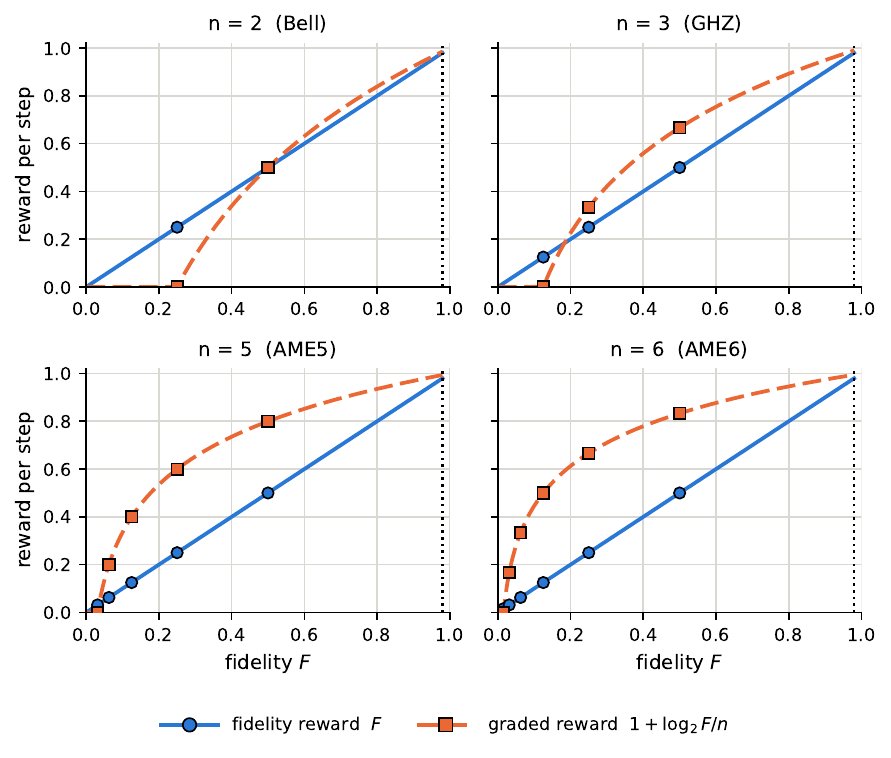}
    \label{fig:reward-shapes}
\end{figure}

\paragraph{Reward.} After each step the circuit $C_{t+1}=\mathrm{Dec}(z_{t+1})$ is simulated and its fidelity with the target is computed. The fidelity is the one defined in Eq.~\eqref{eq:fidelity}, $F(C)=|\braket{\psi^\star|\psi_C}|^2$, where $\ket{\psi_C}$ is the state obtained by applying $C$ to $\ket{0}^{\otimes n}$. We compute it exactly: the circuit is simulated as a state vector, without noise and without sampling, and the overlap of the result with the state vector of the target is taken. The only exception is the noise ablation of Sec.~\ref{sec:results-noise}, where the circuit is simulated with depolarising noise and the fidelity is that of the resulting mixed state. For the two-qubit level, where any of the four Bell states of App.~\ref{app:bell} is accepted, $F$ is the largest of the four fidelities. The circuit is then rewarded with the graded log-fidelity reward
\begin{equation}
    r_{t+1}=\begin{cases}
        50\,F(C_{t+1}) & \text{if } F(C_{t+1})\ge\xi,\\[2pt]
        \max\bigl(0,\;1+\tfrac1n\log_2F(C_{t+1})\bigr) & \text{otherwise,}
    \end{cases}
    \label{eq:log-reward}
\end{equation}
with acceptance threshold $\xi=0.98$.

The factor $50$ in the first case is the bonus for reaching the target. Below the threshold the reward never exceeds $1$ per step, so an episode of at most ten moves that never reaches the target collects at most $10$ in total, whereas a single step on a solution pays about $50$. Reaching a solution is therefore always worth more than any time spent on partial solutions, and because an episode continues after a solution is found, the agent is paid the bonus again for every further step on which it stays on one. The value $50$ is the one used by the fidelity reward of the gate-based implementation we build on~\cite{kundu2024kanqas}; we kept it so that both agents receive the same bonus, and did not tune it. Appendix~\ref{app:bonus} shows that the result is not sensitive to this value: bonuses of $10$, $50$ and $200$ lead to the same final success rates at $n=5$, whereas a bonus of $1$, for which a solution is worth little more than a partial solution, finds the target equally often but learns it markedly worse.

Below the threshold the reward is graded. For stabiliser circuits and targets the fidelity can only take the values $0$ and $2^{-(n-m)}$, where $m$ is the number of independent stabilisers the prepared state shares with the target, so the reward equals $m/n$: it increases in equal steps of $1/n$ for every additional correct stabiliser. Figure~\ref{fig:reward-shapes} compares this reward with the commonly used reward $r=F$~\cite{kundu2024kanqas}, which gives almost no signal on the way: a six-qubit circuit that already shares four of six stabilisers with the target has $F=0.25$, against $0.016$ with none.

\paragraph{Episodes.} An episode consists of $L$ latent moves ($L=6$ for $n\le3$ and $L=10$ for $n\ge5$) and is not terminated when a solution is reached, so the agent is rewarded for staying in regions that decode to solutions. With probability $1/2$ an episode starts at the latent mean $\mu_\phi(C)$ of a seed circuit $C$, chosen as described in Sec.~\ref{sec:method-refresh}, and otherwise at a sample of the prior, $z_0\sim\mathcal{N}(0,I_d)$.

\paragraph{Agent.} The agent is the DDQN of Sec.~\ref{sec:prelim-ddqn}. Its Q-network is a multilayer perceptron (MLP) with three hidden layers of $30$ neurons and ReLU activations. We take the training hyperparameters from KANQAS~\cite{kundu2024kanqas}, but not its network: KANQAS replaces the MLP by a Kolmogorov-Arnold network, whereas we use the MLP throughout, for the latent agent and for the gate-based baseline alike. All values are listed in App.~\ref{app:hyperparameters}. The agent's input and output sizes depend only on $d$, so the same network can be carried from block to block and, in principle, between qubit counts.

\FloatBarrier  
\subsection{Periodic refresh of the latent space}
\label{sec:method-refresh}

In Sec.~\ref{sec:method-mdp} the VAE is kept fixed while the agent acts, so the agent can only find circuits that the latent space already contains. This section removes that limitation by closing the loop between the two components. The circuits that the agent decodes during its episodes become new training data for the circuit VAE of Sec.~\ref{sec:method-vae}, and the refreshed VAE defines the latent space in which the agent's next episodes take place. The refresh reuses what the previous sections introduced: the circuits are written in the canonical gate order of Sec.~\ref{sec:method-encoding}, the VAE is trained with the same objective, free bits and word dropout, and the agent's network is carried over unchanged, which is possible because its input and output sizes depend only on the latent dimension.

The loop is needed because a VAE trained on random circuits represents the target poorly: for the six-qubit AME state, none of $40000$ random circuits from $\mathcal{G}$ prepares it, so the target is not in the training data and need not lie anywhere in the latent space. Following weighted retraining in molecular design~\cite{tripp2020weighted}, we therefore let the latent space follow the agent. Training is split into blocks of $M$ episodes, with $M=1000$ in all our experiments. During a block the VAE is kept fixed, and every circuit the agent decodes that is a solution, or a \emph{partial solution} at least as good as the best circuit found so far, is added to an accumulated pool. After the block, the VAE is fine-tuned for $15$ epochs on $2000$ random circuits together with the pool, in which each pool circuit is repeated $25$ times (circuits much shorter than the longest found so far only once) and each solution $\min\bigl(300,\max(25,\lfloor3000/N_{\mathrm{sol}}\rfloor)\bigr)$ times, where $N_{\mathrm{sol}}$ is the number of distinct solutions; the cap keeps the training set bounded when many distinct solutions exist, as for Bell states. The target itself is never added: every circuit in the pool was produced by the agent. The next block starts from the agent's network and exploration rate at the end of the previous block, and seeds its episodes from the best partial solutions in the pool, so that the agent starts where the search has made most progress.

Refreshing changes the latent space in a way the agent can exploit. Before the first solution is found, the region around the best partial solutions is sharpened; once a solution has been found, it and the circuits near it are amplified, the region of the latent space that decodes to solutions grows, and the agent learns to reach it from most starting points, as the results of Sec.~\ref{sec:results-ladder} show.

\subsection{Transfer across qubit counts}
\label{sec:method-transfer}

The refresh of Sec.~\ref{sec:method-refresh} improves the latent space for one qubit count, using the circuits the agent finds there. Transfer applies the same idea across qubit counts: the solutions collected in the accumulated pool at a smaller qubit count $n'$ become the starting material for the latent space at $n$ qubits. This is the step for which the qubit-count-independent encoding of Sec.~\ref{sec:method-encoding} was designed, since a circuit on $n'$ qubits can be rewritten as a circuit on $n$ qubits without changing the form of its tokens, and the agent of Sec.~\ref{sec:method-mdp} needs no modification because its state and actions do not depend on $n$.

Concretely, the latent space for $n$ qubits is built from the circuits solved at the previous, smaller qubit count $n'$. We take the solutions in the accumulated pool of the best $n'$-qubit run and multiply them by operations that do not depend on the new target: (i) injective relabelings of the $n'$ qubits into the $n$-qubit register, (ii) random reorderings of commuting gates, and (iii) \emph{growth}, which extends a relabelled circuit to the unused qubits by adding graph vertices, each with a Hadamard and $\mathrm{CZ}$ gates to between one and four randomly chosen qubits of the circuit. Up to $200$ relabelled and $200$ grown variants, in canonical order and no longer than the gate budget of the new target, form the \emph{transfer pool}. A new $n$-qubit VAE is trained on the transfer pool, repeated $25$ times, and $2000$ random circuits, and the agent's episodes are initially seeded from the transfer pool's best circuits. Nothing about the new target is used to build the pool; we only check it afterwards and report in Sec.~\ref{sec:setup} how close its best circuits already are.

What is carried from one qubit count to the next is therefore the representation, in the form of a few hundred circuits from which the next latent space is grown, not the agent's network or its replay buffer. Because the agent's input and output sizes are independent of $n$, its network could be carried over as well; we leave this to future work and train a new agent at every qubit count.

\section{Experimental setup}
\label{sec:setup}

\paragraph{Benchmark problems.}
We use the ladder of AME targets of Sec.~\ref{sec:prelim-quantum}: the Bell states for $n=2$, where a circuit is accepted if it prepares any one of the four Bell states defined in App.~\ref{app:bell}, the GHZ state for $n=3$, and the AME states for $n=5$ and $n=6$, the five-qubit ring graph state and the six-qubit graph state of Ref.~\cite{paraskevopoulos2025spin}. For qubits, AME$(2,2)$ and AME$(3,2)$ are the Bell and GHZ states up to local unitaries, so the ladder consists of AME states throughout. The minimal circuits over $\mathcal{G}$ have $2$, $3$, $10$ and $15$ gates. The gate budget $T_{\max}$, which is both the maximum circuit length the VAE represents and the gate-based agent's number of layers, is $6$ for $n\le3$, $15$ for $n=5$ and $20$ for $n=6$. It must leave room above the minimal circuit: at $T_{\max}=10$ for $n=5$, every circuit the agent reached that shared four of five stabilisers with the target already used all ten gates, so that no small change led to the target, and no seed solved it. The levels are run in order $2\to3\to5\to6$, each transfer pool being built from the run with the most solved episodes at the previous level. No circuit in any transfer pool prepares the next target. At $n=5$ the best pool circuits share three of five stabilisers with AME5, and no sequence of up to three added gates turns any of them into the target; at $n=6$ one grown circuit shares four of six stabilisers with AME6 and is a single $\mathrm{CZ}$ gate away from it. We keep it, since it was produced without knowledge of the target, but note that it makes the six-qubit level easier to reach from the transfer pool.
For completeness we also run an $n=4$ level, the four-qubit GHZ state (minimal circuit $4$ gates, $T_{\max}=8$, $L=6$), with its transfer pool grown from the same $n=3$ run. No qubit AME state exists for $n=4$, so this level is not part of the ladder and is not used to build the $n=5$ and $n=6$ pools; it shows that transfer works at every qubit count, not only between the AME levels.

\paragraph{Training budget.}
Every run has the same budget of $M=1000$ episodes per block: $10$ blocks for $n\le3$ and $15$ blocks for $n\ge5$, that is, $10\,000$ and $15\,000$ episodes. Each level is run with ten seeds. Deep reinforcement learning results can vary strongly from one random seed to the next, and ten or more seeds are recommended for a thorough study~\cite{achiam2018spinningup}.

\paragraph{Baselines.}
\begin{itemize}
    \item \textbf{Gate-based DDQN.} The standard RL-QAS agent of Sec.~\ref{sec:prelim-qas}, in the KANQAS implementation~\cite{kundu2024kanqas}, the same gate-based RL-QAS implementation as in Refs.~\cite{ostaszewski2021rl,kuo2021qasdrl,fosel2021circuitopt,kundu2024vqsd,kundu2024rlqasthesis}, with an MLP as its Q-network and identical agent hyperparameters, building each circuit gate by gate from an empty circuit. It receives the same number of episodes ($15\,000$ for $n\ge5$) and a gate budget equal to $T_{\max}$. We run it with the original fidelity reward ($10\,000$ episodes) and with the graded reward of Eq.~\eqref{eq:log-reward}, and with $\mathrm{CX}$ or $\mathrm{CZ}$ as its two-qubit gate, five seeds each.
    \item \textbf{Random latent moves.} To separate what the trained policy contributes from what the refreshed latent space contributes, we re-evaluate every block of the $n=5,6$ runs with $\epsilon=1$. This control stays entirely within the RL framework: it uses the same environment as the trained agent, with the same VAE, episode starts, actions and episode length, the same reward is computed after every move, and an episode counts as solved by the same criterion. The only difference is that every move is drawn uniformly at random instead of being chosen by the policy, and that the reward received is not used for learning.
    \item \textbf{Ablations.} \method{} without transfer; the gate-based agent with the latent agent's transfer pool, and with a ladder of its own. They are described with their results in Sec.~\ref{sec:results-ablations}.
\end{itemize}

\paragraph{Noise.}
All experiments are noiseless, except for a noise ablation in which a depolarising channel is applied after every gate, with error probability $p_1=0.001$ on the qubit of a one-qubit gate and $p_2=0.01$ on the two qubits of a two-qubit gate. The circuit then prepares a mixed state $\rho$, and every fidelity used during training, in the reward of Eq.~\eqref{eq:log-reward} as well as in the selection of circuits for the refresh and for transfer, is the noisy fidelity $F=\langle\psi|\rho|\psi\rangle$ with the ideal target $|\psi\rangle$. Since even an exact circuit no longer reaches $F=1$, the acceptance threshold is lowered to $\xi=0.75$. We run two independent settings, five seeds each. In the first, the levels $n=2$ and $n=3$ are noiseless and noise is present at $n=5$ and $n=6$. In the second, noise is present at every level. Appendix~\ref{app:noise} gives the details.

\paragraph{Metrics.}
For each level we report (i) the number of seeds that prepare the target at least once, and the training block of $M=1000$ episodes in which the first solution is found; (ii) the \emph{success rate}: after each block the trained agent is evaluated on $500$ episodes without exploration ($\epsilon=0$) and without further training, and the success rate is the fraction of these episodes in which it prepares the target, $F\ge\xi$. It measures what the agent has learned rather than its exploration; (iii) the fraction of prior samples $z\sim\mathcal{N}(0,I)$ that decode to a solution, a property of the latent space alone; and (iv) the number of distinct solution circuits found. To characterise the latent space we report, for a population of circuits encoded to their latent means, the number of \emph{active} dimensions, those with variance above $0.01$~\cite{burda2016iwae}, and the participation ratio $(\sum_k\lambda_k)^2/\sum_k\lambda_k^2$ of the covariance eigenvalues $\lambda_k$, the effective number of dimensions the population spreads over.

\paragraph{Implementation details.}
All circuits are simulated as state vectors with Qiskit~\cite{javadiabhari2024qiskit}. The VAE has embedding size $16$, token width $64$, hidden size $64$ and $d=8$; the initial VAEs are trained for $80$ epochs with ADAM (learning rate $2\cdot10^{-3}$, batch size $64$) and each refresh fine-tunes for $15$ epochs at learning rate $5\cdot10^{-4}$. Agent hyperparameters are listed in App.~\ref{app:hyperparameters}. All runs were performed on a single laptop (Intel Core Ultra 7 155H CPU, 16\,GB RAM, no GPU), with the seeds of a level run in parallel. One seed takes $72$, $45$, $58$, $86$ and $93$ minutes of wall-clock time at $n=2$, $3$, $4$, $5$ and $6$, of which $35$-$54\%$ is spent refreshing the VAE and the rest on RL episodes and greedy evaluation; building a transfer pool and training the initial VAE of a level takes a few minutes.

\section{Results}
\label{sec:results}

We present the results in four steps. Section~\ref{sec:results-ladder} gives the main result, the performance of \method{} along the ladder of targets from Bell pairs to the six-qubit AME state, measured by how many seeds find each target and how reliably the trained policy prepares it. It also separates the contribution of the trained policy from that of the refreshed latent space, using the random-move control, and describes the set of solution circuits a run produces. Section~\ref{sec:results-noise} repeats the ladder with depolarising noise on every gate. Section~\ref{sec:results-baseline} compares \method{} with its baselines: a gate-based DDQN that receives the same number of episodes, and variants that separate the effect of transfer from the effect of the latent representation. Finally, Sec.~\ref{sec:results-representation} analyses the latent space itself: how many of its dimensions the VAE uses at each qubit count, and how the periodic refresh changes this.

\subsection{Transfer along the AME ladder}
\label{sec:results-ladder}

\begin{figure}[t!]
    \centering
    \caption{\textbf{Success rate along the AME ladder (ten seeds per level)}. Solid blue: \method{}, the success rate of the trained agent after each training block. Dashed orange: a control in which the agent makes uniformly random moves, in the same latent space and from the same episode starts ($n=5,6$). Dotted green: the latent space alone, the fraction of latent points drawn at random from the prior that decode to a target-preparing circuit. Lines are medians and bands the range over the seeds that found the target.}
    \includegraphics[width=0.8\linewidth]{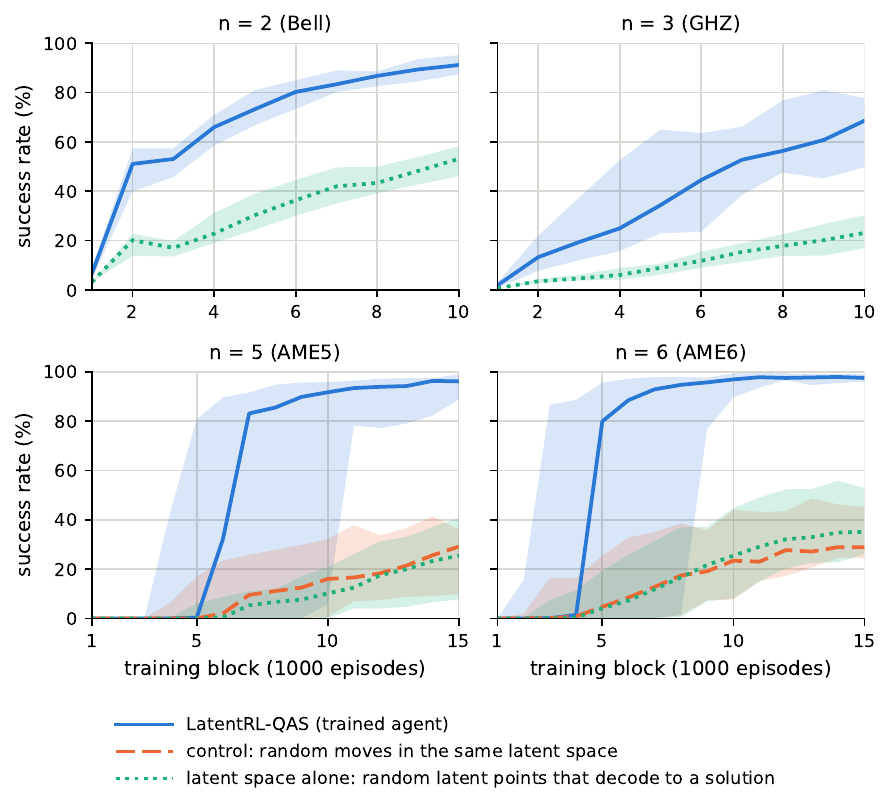}
    \label{fig:ladder}
\end{figure}

Figure~\ref{fig:ladder} and Table~\ref{tab:ladder} summarise how \method{} performs along the ladder. For each level we report two things: whether the target is \emph{found}, and how well it is then \emph{learned}, measured by the success rate defined in Sec.~\ref{sec:setup}. Time is counted in training blocks. A \emph{block} is $M=1000$ consecutive training episodes, after which the latent space is refreshed, as defined in Sec.~\ref{sec:method-refresh}; it is unrelated to the size of the circuit.

\paragraph{Finding the targets.} The Bell and GHZ states are prepared by all ten seeds in the first block, that is, within the first $1000$ episodes. For AME5, all ten seeds find the target, in blocks $3$ to $9$ (median $5$); AME6 is also found in all ten seeds, in blocks $1$ to $7$ (median $3$).

\paragraph{Learning the targets.} Once a seed has found a target, its success rate rises within one or two blocks, as the solid blue curves of Fig.~\ref{fig:ladder} show. After the last block it is $88.8$-$99.0\%$ for AME5 and $96.0$-$99.0\%$ for AME6. The two small levels are learned more gradually. The Bell level ends at $87.4$-$95.2\%$ after ten blocks and the GHZ level at $49.8$-$77.8\%$. The GHZ level is still improving at that point: extending it to $15$ blocks raises all ten seeds to $61.4$-$88.2\%$.

\paragraph{The additional four-qubit level.} This level is run for completeness and is not part of the ladder, as explained in Sec.~\ref{sec:setup}. It gives the same picture as the two- and three-qubit levels. None of the circuits in its transfer pool prepares the four-qubit GHZ state, so the agent has to find it. All ten seeds do so in the first block, and the success rate then rises steadily to $67.4$-$90.8\%$ after ten blocks.

\begin{table}[h!]
    \centering
    \small
    \caption{\small\textbf{Results per level of transfer, ten seeds each}. First solving block: the training block of $1000$ episodes in which the target is first prepared. Greedy solve rate after the last block, as min-max over the seeds that found the target. $^\ast$Completion level, not used to build the $n=5,6$ pools.}
    \label{tab:ladder}
    \begin{tabular}{llrrrr}
        \toprule
        $n$ & Target & Episodes & Seeds solved & First solving block & Final success rate \\
        \midrule
        2 & Bell & 10\,000 & 10/10 & 1 & 87.4-95.2\% \\
        3 & GHZ  & 10\,000 & 10/10 & 1 & 49.8-77.8\% \\
        3 & GHZ (15 blocks) & 15\,000 & 10/10 & 1 & 61.4-88.2\% \\
        4$^\ast$ & GHZ(4) & 10\,000 & 10/10 & 1 & 67.4-90.8\% \\
        5 & AME5 & 15\,000 & 10/10 & 3-9 & 88.8-99.0\% \\
        6 & AME6 & 15\,000 & 10/10 & 1-7 & 96.0-99.0\% \\
        \bottomrule
    \end{tabular}
\end{table}

\paragraph{What the trained policy adds.} A high success rate could in principle come from the latent space and not from the agent, because refreshing the latent space with the agent's discoveries enlarges the region that decodes to solutions. The dotted green curves of Fig.~\ref{fig:ladder} show this effect: by the last block, $8$-$41\%$ of latent points drawn at random from the prior decode to an AME5 circuit and $26$-$53\%$ to an AME6 circuit, against none before the first solution. To separate the two contributions we use the random-move control of Sec.~\ref{sec:setup}, shown as the dashed orange curves. In the same refreshed spaces and from the same episode starts, uniformly random moves solve $10$-$36\%$ of AME5 episodes and $24$-$45\%$ of AME6 episodes in the last block. The trained policy solves $89$-$99\%$ and $96$-$99\%$. The policy thus adds $54$ to $82$ percentage points over the latent space alone. Its advantage appears as soon as the target is found: within one or two blocks of the first solution the policy solves most episodes, while random moves in the same space solve at most $17\%$.

\paragraph{The circuits a run produces.} The outcome of a run is not a single circuit but a set of them. Each run that found its target produced several hundred distinct solution circuits ($153$-$614$ for AME5 and $458$-$645$ for AME6). Most of this variety is in the order of the gates and in the direction of the $\mathrm{CX}$ gates; structurally, the solutions fall into $27$ (AME5) and $22$ (AME6) distinct sets of coupled qubit pairs across all seeds. All AME6 solutions have the same composition, six $H$, two $\mathrm{CX}$ and nine $\mathrm{CZ}$ gates, $17$ in total against $15$ for the minimal circuit; the AME5 solutions have $12$-$15$ gates, with $5$ to $10$ two-qubit gates, none of the $2998$ is minimal, and only $4$ use $H$ and $\mathrm{CZ}$ gates alone, as the minimal circuit does. The refreshed latent space thus concentrates on one structural family of circuits per target and offers some twenty different qubit couplings within it, from which a circuit can be chosen according to the hardware's connectivity. Combined with a policy that reaches a solution from almost any starting point, it acts as an on-demand generator of circuits for the target state.

\FloatBarrier  
\subsection{Robustness to gate noise}
\label{sec:results-noise}

The results so far were obtained with noiseless simulation. To test whether the method depends on exact fidelities, we repeat the ladder with the depolarising noise of Sec.~\ref{sec:setup} in two settings: noise at the two largest levels only, and noise at every level. Table~\ref{tab:noise} compares both with the noiseless ladder.

\method{} is not affected by noise of this strength. In both settings every target is found, and the trained policies prepare it at rates comparable to the noiseless ones. With noise at $n=5,6$ only, AME5 is found in four of five seeds and AME6 in all five, with final success rates of $94.8$-$97.6\%$ and $96.4$-$98.8\%$. With noise at every level, all five seeds find every target. The final rates at $n=6$ are then slightly lower, $89.6$-$97.2\%$, and AME5 is first found in blocks $3$ to $10$, a wider spread than in the other setting.

Noise also does not lead the agent to accept wrong circuits. Every circuit that passed the noisy threshold, in either setting and at every level, prepares the target exactly when simulated without noise. The agent therefore does not exploit the lower threshold; it finds the same kind of solution as before, evaluated through a noisy channel. The noisy fidelity of these solutions is $0.89$-$0.94$ at $n=5,6$, far above the threshold.

\begin{table}[h!]
    \centering
    \small
    \caption{\small \textbf{The transfer ladder under depolarising noise ($p_1=0.001$, $p_2=0.01$)}. Seeds that found the target, and final success rate as min-max over those seeds. The noiseless ladder has ten seeds, the noisy settings five.}
    \label{tab:noise}
    \begin{tabular}{llcccccc}
        \toprule
        & & \multicolumn{2}{c}{Noiseless} & \multicolumn{2}{c}{Noise at $n=5,6$} & \multicolumn{2}{c}{Noise at every level} \\
        \cmidrule(lr){3-4}\cmidrule(lr){5-6}\cmidrule(lr){7-8}
        $n$ & Target & Found & Learned & Found & Learned & Found & Learned \\
        \midrule
        2 & Bell & 10/10 & 87.4-95.2\% & \multicolumn{2}{c}{noiseless} & 5/5 & 85.6-96.4\% \\
        3 & GHZ  & 10/10 & 49.8-77.8\% & \multicolumn{2}{c}{noiseless} & 5/5 & 53.2-79.0\% \\
        5 & AME5 & 10/10  & 88.8-99.0\% & 4/5 & 94.8-97.6\% & 5/5 & 85.0-100.0\% \\
        6 & AME6 & 10/10 & 96.0-99.0\% & 5/5 & 96.4-98.8\% & 5/5 & 89.6-97.2\% \\
        \bottomrule
    \end{tabular}
\end{table}

\FloatBarrier  
\subsection{Comparison with baselines}
\label{sec:results-baseline}
\label{sec:results-ablations}

Table~\ref{tab:ablations} compares \method{} with its baselines on the two AME targets, five seeds each (ten for the full method) with the budgets of Sec.~\ref{sec:setup}. The latent agent differs from the standard gate-based agent in two ways: it acts in a learned representation, and it starts each level from circuits transferred from the previous one. We therefore compare it first with the gate-based agent trained from scratch, and then with three baselines that separate the two differences. Throughout, we distinguish \emph{finding} a target, preparing it at least once, from \emph{learning} it, preparing it reliably.

\paragraph{Gate-based DDQN from scratch.} With the same number of episodes and identical agent hyperparameters, the gate-based DDQN does not prepare either AME target. With $\mathrm{CX}$ as its two-qubit gate and the fidelity reward ($10\,000$ episodes), none of five seeds solves AME5 or AME6, the best circuits reaching fidelities of $0.5$ and $0.125$; with the graded reward of Eq.~\eqref{eq:log-reward}, none of five seeds solves either target, the best circuits sharing three or four of five stabilisers with AME5 and four of six with AME6. Giving it $\mathrm{CZ}$, the gate the AME graph states are built from, does not change this: again none of five seeds solves AME5 or AME6. Over the three variants, that is $0$ of $30$ runs. \method{}, with the same budget, solves AME5  and AME6 in all ten seeds, as listed in Table~\ref{tab:ladder}. At the scale of Table~\ref{tab:sizes}, it does so with a network of $2626$ parameters against $25\,785$ and $43\,350$ for the gate-based agent at $n=5$ and $n=6$. On the small targets the gate-based agent is competitive, as expected: with $\mathrm{CX}$, the graded reward and $10\,000$ episodes it prepares the GHZ state in all five seeds, solving $48$-$50\%$ of its training episodes. The advantage of the latent representation therefore appears where the gate-based state and action spaces become large, as quantified in Table~\ref{tab:sizes}.

\paragraph{Transfer is necessary.} Without transfer, the VAE for $n=5$ or $6$ is trained on random circuits only and episodes start from prior samples, with everything else unchanged. The agent then finds AME5 in one of five seeds, in block $13$ of $15$, reaching a success rate of $48.8\%$ by the end, and AME6 in none. The transfer pool supplies the partial solutions from which the target becomes reachable; random circuits essentially never contain them.

\paragraph{Transfer is not sufficient.} Conversely, we give the gate-based agent the transfer pool of \method{}. Using $\mathrm{CZ}$ and the same best-partial seeding, half of its episodes start by replaying one of the best pool circuits, and the pool is extended with its own best partial solutions. At $n=5$ it finds AME5 in two of five seeds, solving $14$-$35\%$ of its training episodes towards the end. At $n=6$, where one pool circuit is a single $\mathrm{CZ}$ gate away from AME6, as noted in Sec.~\ref{sec:setup}, all five seeds reach AME6, but only $1$ to $8$ times each in $15\,000$ episodes, and never more than once in the last $1000$: it stumbles on the target from a nearby start without learning to reach it. \method{}, given the same pool, ends at $96$-$99\%$.

\paragraph{The gate-based agent with its own ladder.} The pool above was built from the discoveries of \method{}. To give the gate-based agent the same kind of curriculum without borrowing anything from the latent method, we let it build its own ladder: GHZ from scratch at $n=3$, then AME5 from a pool grown in the same way from its own GHZ solutions, then AME6 from a pool grown from its own AME5 solutions. It finds AME5 in all five seeds, but with very different success, solving $1$, $36$, $52$, $1115$ and $1134$ of $15\,000$ training episodes, and at most $16\%$ of the last $1000$. Its AME6 pool, whose best circuits share three of six stabilisers with the target, is weaker than that of \method{}: it finds AME6 in two of five seeds, $5$ and $2$ times, and never learns it. A curriculum thus makes the larger targets \emph{findable} for the gate-based agent, but it does not make them \emph{learnable}: its state and action spaces grow with $n$, while the latent agent's do not.

\begin{table}[h!]
    \centering
    \small
    \caption{\small\textbf{Comparison of \method{} with its baselines}. We use five seeds for each method and ten for the full method. We see that seeds that prepare the target at least once. Learned: final success rate for \method{}, and training solve rate over the last $1000$ episodes for the gate-based agent (with $\epsilon\ge0.05$), over the seeds that found the target.}
    \label{tab:ablations}
    \begin{tabular}{lcccc}
        \toprule
        & \multicolumn{2}{c}{AME5 ($n=5$)} & \multicolumn{2}{c}{AME6 ($n=6$)} \\
        \cmidrule(lr){2-3}\cmidrule(lr){4-5}
        Method & Found & Learned & Found & Learned \\
        \midrule
        \textbf{\method{} (full method)} & \textbf{10/10} & \textbf{89-99\%} & \textbf{10/10} & \textbf{96-99\%} \\
        \method{}, no transfer & 1/5 & 49\%     & 0/5 & --- \\
        Gate DDQN, from scratch (3 variants)& 0/15 & ---     & 0/15 & --- \\
        Gate DDQN, pool of \method{} & 2/5 & 14-35\% & 5/5 & $\le0.1\%$ \\
        Gate DDQN, own ladder               & 5/5 & 0-16\%  & 2/5 & $\le0.1\%$ \\
        \bottomrule
    \end{tabular}
\end{table}

\FloatBarrier  
\subsection{Analysis of the latent space}
\label{sec:results-representation}

\begin{figure}[t!]
    \centering
    \small
    \caption{\small\textbf{Illustration of how the latent dimensions are used}. For each level (rows) and for the VAE the level starts from (left) and the last refreshed VAE (right), bars show the fraction of variance along each principal component of the latent means of $2000$ random circuits (plain blue bars) and of the distinct target-preparing circuits found by the best run (hatched orange bars); each panel states the two participation ratios. The top-left VAE is the original two-qubit model trained without free bits.}
    \includegraphics[width=0.8\linewidth]{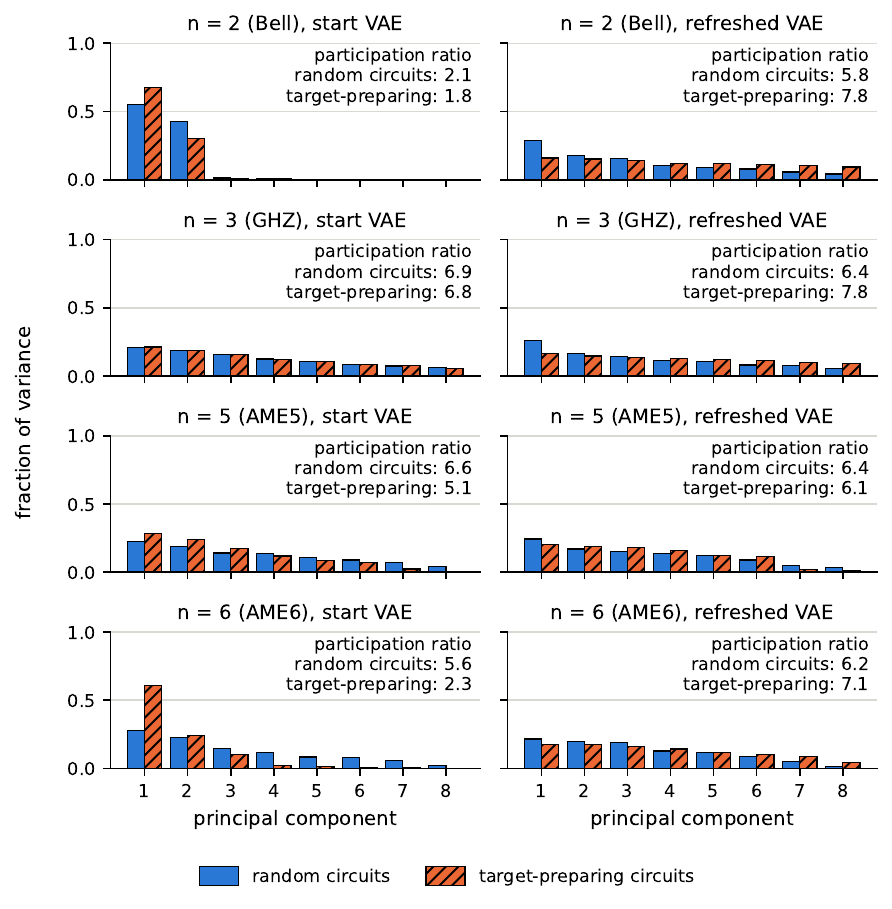}
    \label{fig:latent-geometry}
\end{figure}

Our first experiments, at $n=2$ with a VAE trained without free bits, already showed that the latent space is organised rather than filled at random: the Bell-state circuits the agent found occupied an effective $2.5$ of the $8$ dimensions, with three principal components explaining $97\%$ of their variance, and the four Bell states lay almost exactly on a line. The VAE also adapts how many dimensions it uses to what the data requires. The two-qubit circuit space is small, and the two-qubit VAE the ladder starts from encodes it in $4$ active dimensions, with random circuits spreading over an effective $2.1$ of them, as the top-left panel of Fig.~\ref{fig:latent-geometry} shows; the remaining dimensions are left free. This compact space is fully navigable: starting from it, the agent learns to prepare Bell states in $87$-$95\%$ of evaluation episodes, as reported in Sec.~\ref{sec:results-ladder}.

For larger circuits this behaviour has to be controlled. Without free bits, the six-qubit VAE collapsed entirely, as described in Sec.~\ref{sec:method-vae}; with free bits and word dropout every dimension is kept in use, as Fig.~\ref{fig:latent-geometry} shows. At every level, all $8$ dimensions are active and random circuits spread over an effective $5.6$ to $6.9$ dimensions. Two observations stand out. First, \emph{the dimension the representation needs does not grow with the number of qubits}: the same eight dimensions hold the two-qubit and the six-qubit circuit populations with essentially the same effective spread, consistent with the qubit-count-independent encoding of Sec.~\ref{sec:method-encoding}. Second, \emph{refresh reallocates the latent space to the solutions}. Before refresh, the circuits that prepare the target occupy a low-dimensional part of the space, with effective dimension $1.8$ for Bell states and $2.3$ for AME6. After refresh they span $7.8$ and $7.1$ dimensions: the latent space devotes most of its capacity to the hundreds of distinct solutions the agent has found, which are analysed in Sec.~\ref{sec:results-ladder}, rather than to a single point.

The VAEs built by transfer are also faithful and diverse. Decoding $1000$ prior samples gives $925$, $992$ and $973$ distinct circuits at $n=3$, $5$ and $6$. At $n=5$ and $6$ none of them prepares the target, so before the agent finds it the target cannot be reached by sampling the space.

\FloatBarrier  
\section{Discussion}
\label{sec:discussion}

We introduced \textbf{\method{}}, \textit{an effective way to transfer knowledge across qubit counts through a latent space representation of circuits, within the RL framework}. In our experiments, what is carried from one system size to the next is this representation, and not the policy. Our findings can be summarised in the following two points, each of which rests on one of the ablations of Sec.~\ref{sec:results-ablations}. (1) Transferred circuits make a larger target \emph{findable}. This follows from the ablation that removes transfer from \method{}: without transferred circuits it does not find AME6, and with them it does in every seed. (2) The latent representation makes a found target \emph{learnable}. This follows from the ablations that give the gate-based agent the same transferred circuits, or a curriculum of its own: it then at best stumbles on AME6 a few times, whereas the latent agent learns to reach it from almost any starting point within one or two blocks of the first solution. We attribute this to the size of the problem the agent faces. The gate-based agent's observation and action spaces grow with $n$, as listed in Table~\ref{tab:sizes}, and a solution found from one starting circuit tells it little about others. The latent agent sees $8$ numbers and chooses among $16$ moves at every qubit count, and refreshing the latent space moves the solutions it has found closer to every starting point. To achieve this, the method relies on two components: a state representation for the RL agent whose overhead is constant in the number of qubits, and the periodic refresh of the latent space. We discuss them in turn. (1) \textbf{Constant qubit overhead circuit encoding} which keeps the state of the agent from growing with the system. The latent space uses all eight of its dimensions at every qubit count, with an effective dimension of $5.6$-$6.9$ for random circuits from two to six qubits, and the VAE grows by $2\%$ from $n=2$ to $n=10$. Nothing in the encoding, the VAE or the agent has to be redesigned when qubits are added; what changes is the content of the latent space, not its size. (2) \textbf{Periodic refresh of the latent space} which retrains the VAE on the circuits the agent discovers. It turns a solution found once into a solution that is reached reliably: once a run has found the target, its success rate rises within one or two blocks. It also changes what a run produces. The product of a run is not a single circuit but a refreshed latent space together with a policy that navigates it. The space contains hundreds of distinct solutions per target, in about twenty different qubit couplings, and the policy reaches one from almost any starting point. Whenever a circuit for the target is needed, for instance for a device with a particular connectivity, it can be drawn from this space rather than searched for again.

\section{Conclusion}
\label{sec:conclusion}

We introduced \method{}, a way to transfer what an RL agent learns in quantum architecture search from one system size to the next, by transferring a learned representation of circuits rather than the agent or its experience. A circuit VAE built on an operation-level encoding maps circuits of any qubit count to a latent space of fixed dimension; a DDQN agent of $2626$ parameters navigates this space; the space is refreshed with the agent's discoveries; and the space for $n$ qubits is grown from the solutions found at a smaller $n$. On a ladder of absolutely maximally entangled targets from two to six qubits, \method{} finds AME5 and AME6 in all ten seeds, without the target ever being supplied, and learns to prepare them in $89$-$99\%$ and $96$-$99\%$ of evaluation episodes. A gate-based DDQN with the same budget never solves either target from scratch, and even when given the same transferred circuits or its own curriculum it does not learn to prepare AME6. The refreshed latent space ends up holding hundreds of distinct circuits for each target, so that together with the policy it acts as a generator of target-preparing circuits. Because neither the representation nor the agent grows appreciably with the number of qubits, we expect the approach to extend to the most uniform graph states on seven to ten qubits, which we leave to future work.

\section{Limitations and future work}
\label{sec:limitations}

\paragraph{Limitations.} (1) At $n=6$ the transfer pool, grown without knowledge of the target, happens to contain a circuit one gate away from AME6. This makes AME6 easy to \emph{find} for every method started from the pool, so the six-qubit level mainly tests learning; the five-qubit level, whose pool circuits are more than three gates away from the target, is the cleaner test of discovery. (2) Once a solution has been found, refreshing amplifies it. Part of the late success rate therefore comes from the space itself. (3) The solutions are not minimal. Most AME6 solutions use $17$ gates against $15$, and none of the $2998$ AME5 solutions is minimal; the reward does not penalise length. (4) Finally our gate set is discrete and Clifford, our targets are graph states, and a new agent is trained at every qubit count. The noise we tested is simulated depolarising noise of one strength; we did not test device-specific noise models or real hardware.

\paragraph{Future work.} The most direct extension is to more qubits. AME states do not exist beyond $n=6$, but the ladder continues naturally with the most uniform graph states, those whose every subset of $k$ qubits is maximally mixed for the largest possible $k$. We have identified and verified such targets for $n=7$ to $10$: a $2$-uniform state on the seven-qubit ring ($14$ gates; an exhaustive search shows that no $3$-uniform graph state exists on seven qubits), and $3$-uniform states for $n=8$ (the cube graph, $20$ gates), $n=9$ ($23$ gates) and $n=10$ ($25$ gates). Since the VAE and the agent hardly change with $n$, as Table~\ref{tab:sizes} shows, these experiments require no change to the method. Other directions are to carry the agent's network, not only the latent space, from one qubit count to the next; to use the angle field of the token of Eq.~\eqref{eq:token} for parameterised gates and variational tasks; to add a length or two-qubit-gate penalty that favours minimal and hardware-native circuits; and to train under device-specific noise models and on real hardware.

\paragraph{Beyond spatial proximity: does the LatentRL-QAS learn physics?} Our transfer rests on an assumption about how difficulty scales: that a circuit for a harder target is largely built up from circuits for simpler ones, so that similar circuits map to nearby latent points and a solution at $n$ qubits can be grown from solutions at $n'<n$. This held along our ladder, but it need not hold in general. There may be a level of difficulty at which the next target is not an extension of the previous one but requires a qualitatively different circuit structure. Our own ladder hints at such a point. Beyond $n=6$ no AME state exists, and the most uniform seven-qubit graph state is only $2$-uniform, so the structure of the target class itself changes. How the latent space and the agent behave at such a transition is an open question. Can the refreshed space reorganise itself around the new structure, or does transfer then hinder rather than help? Answering it requires understanding \emph{what} the latent space encodes, beyond which circuits lie close to each other. We are motivated here by SciNet~\cite{iten2020scinet}, which showed that a network whose architecture mirrors a simplified modelling process learns, in its latent representation, the physically relevant parameters of a system, and can thereby yield conceptual insight. In the same spirit, we plan to investigate whether individual latent directions, or the regions and paths the agent prefers, correlate with physical properties of the prepared states: their entanglement across bipartitions, their degree of $k$-uniformity, the number of stabilisers they share with the target, or the structure of the underlying graph. That the four Bell states lie almost on a line, as found in Sec.~\ref{sec:results-representation}, suggests that such structure may be present. If the latent space and the agent's navigation do encode such properties, they could serve not only to find circuits but also as a tool for discovering which properties make a state hard to prepare.

\bibliographystyle{unsrtnat}
\bibliography{references}

@book{nielsen2010quantum,
  title     = {Quantum Computation and Quantum Information},
  author    = {Nielsen, Michael A. and Chuang, Isaac L.},
  edition   = {10th Anniversary},
  publisher = {Cambridge University Press},
  year      = {2010}
}

@article{preskill2018nisq,
  title   = {Quantum Computing in the {NISQ} era and beyond},
  author  = {Preskill, John},
  journal = {Quantum},
  volume  = {2},
  pages   = {79},
  year    = {2018},
  doi     = {10.22331/q-2018-08-06-79}
}

@incollection{greenberger1989ghz,
  title     = {Going Beyond {Bell's} Theorem},
  author    = {Greenberger, Daniel M. and Horne, Michael A. and Zeilinger, Anton},
  booktitle = {Bell's Theorem, Quantum Theory and Conceptions of the Universe},
  editor    = {Kafatos, M.},
  publisher = {Kluwer Academic},
  pages     = {69--72},
  year      = {1989}
}

@article{hein2004graph,
  title   = {Multiparty entanglement in graph states},
  author  = {Hein, Marc and Eisert, Jens and Briegel, Hans J.},
  journal = {Physical Review A},
  volume  = {69},
  pages   = {062311},
  year    = {2004},
  doi     = {10.1103/PhysRevA.69.062311}
}

@article{scott2004multipartite,
  title   = {Multipartite entanglement, quantum-error-correcting codes, and entangling power of quantum evolutions},
  author  = {Scott, A. J.},
  journal = {Physical Review A},
  volume  = {69},
  pages   = {052330},
  year    = {2004},
  doi     = {10.1103/PhysRevA.69.052330}
}

@article{higuchi2000couples,
  title   = {How entangled can two couples get?},
  author  = {Higuchi, A. and Sudbery, A.},
  journal = {Physics Letters A},
  volume  = {273},
  pages   = {213--217},
  year    = {2000},
  doi     = {10.1016/S0375-9601(00)00480-1}
}

@article{huber2017seven,
  title   = {Absolutely Maximally Entangled States of Seven Qubits Do Not Exist},
  author  = {Huber, Felix and G{\"u}hne, Otfried and Siewert, Jens},
  journal = {Physical Review Letters},
  volume  = {118},
  pages   = {200502},
  year    = {2017},
  doi     = {10.1103/PhysRevLett.118.200502}
}

@article{paraskevopoulos2025spin,
  title   = {Near-Term Spin-Qubit Architecture Design via Multipartite Maximally Entangled States},
  author  = {Paraskevopoulos, Nikiforos and Steinberg, Matthew and Undseth, Brennan and Sarkar, Aritra and Vandersypen, Lieven M. K. and Xue, Xiao and Feld, Sebastian},
  journal = {PRX Quantum},
  volume  = {6},
  pages   = {020307},
  year    = {2025},
  doi     = {10.1103/PRXQuantum.6.020307}
}

@inproceedings{ostaszewski2021rl,
  title     = {Reinforcement learning for optimization of variational quantum circuit architectures},
  author    = {Ostaszewski, Mateusz and Trenkwalder, Lea M. and Masarczyk, Wojciech and Scerri, Eleanor and Dunjko, Vedran},
  booktitle = {Advances in Neural Information Processing Systems},
  volume    = {34},
  pages     = {18182--18194},
  year      = {2021}
}

@inproceedings{patel2024crlqas,
  title     = {Curriculum reinforcement learning for quantum architecture search under hardware errors},
  author    = {Patel, Yash J. and Kundu, Akash and Ostaszewski, Mateusz and Bonet-Monroig, Xavier and Dunjko, Vedran and Danaci, Onur},
  booktitle = {The Twelfth International Conference on Learning Representations (ICLR)},
  year      = {2024},
  eprint    = {2402.03500},
  archivePrefix = {arXiv}
}

@article{kundu2025tensorrl,
  title={{TensorRL-QAS}: Reinforcement learning with tensor networks for improved quantum architecture search},
  author={Kundu, Akash and Mangini, Stefano},
  journal={Advances in Neural Information Processing Systems},
  volume={38},
  pages={134032--134066},
  year={2025},
  doi={10.52202/085713-4033}
}

@inproceedings{ikhtiarudin2025benchrl,
  title     = {{BenchRL-QAS}: Benchmarking Reinforcement Learning Algorithms for Quantum Architecture Search},
  author    = {Ikhtiarudin, Azhar and Das, Aditi and Thakkar, Param and Kundu, Akash},
  booktitle = {Proceedings of the AAAI Symposium Series},
  volume    = {7},
  number    = {1},
  pages     = {358--367},
  year      = {2025},
  doi       = {10.1609/aaaiss.v7i1.36906}
}

@inproceedings{kundu2026replay,
  title     = {Replay-buffer engineering for noise-robust quantum circuit optimization},
  author    = {Kundu, Akash and Feld, Sebastian},
  booktitle = {Advances in Neural Information Processing Systems},
  year      = {2026},
  eprint    = {2604.21863},
  archivePrefix = {arXiv}
}

@article{kundu2024kanqas,
  title   = {{KANQAS}: {Kolmogorov-Arnold} network for quantum architecture search},
  author  = {Kundu, Akash and Sarkar, Aritra and Sadhu, Abhishek},
  journal = {EPJ Quantum Technology},
  volume  = {11},
  number  = {1},
  pages   = {76},
  year    = {2024}
}

@article{kundu2026gadgets,
  title   = {Reinforcement learning with learned gadgets to tackle hard quantum problems on real hardware},
  author  = {Kundu, Akash and Sarra, Leopoldo},
  journal = {Communications Physics},
  volume  = {9},
  number  = {1},
  pages   = {44},
  year    = {2026},
  eprint  = {2411.00230},
  archivePrefix = {arXiv}
}

@article{kundu2024rlqasthesis,
  title={Reinforcement learning-assisted quantum architecture search for variational quantum algorithms},
  author={Kundu, Akash},
  journal={arXiv preprint arXiv:2402.13754},
  year={2024}
}

@article{zen2025faulttolerant,
  title   = {Quantum Circuit Discovery for Fault-Tolerant Logical State Preparation with Reinforcement Learning},
  author  = {Zen, Remmy and Olle, Jan and Colmenarez, Luis and Puviani, Matteo and M{\"u}ller, Markus and Marquardt, Florian},
  journal = {Physical Review X},
  volume  = {15},
  number  = {4},
  pages   = {041012},
  year    = {2025},
  doi     = {10.1103/gqpr-dgz7},
  eprint  = {2402.17761},
  archivePrefix = {arXiv}
}

@misc{kuo2021qasdrl,
  title         = {Quantum Architecture Search via Deep Reinforcement Learning},
  author        = {Kuo, En-Jui and Fang, Yao-Lung L. and Chen, Samuel Yen-Chi},
  year          = {2021},
  eprint        = {2104.07715},
  archivePrefix = {arXiv},
  primaryClass  = {quant-ph}
}

@misc{fosel2021circuitopt,
  title         = {Quantum circuit optimization with deep reinforcement learning},
  author        = {F{\"o}sel, Thomas and Niu, Murphy Yuezhen and Marquardt, Florian and Li, Li},
  year          = {2021},
  eprint        = {2103.07585},
  archivePrefix = {arXiv},
  primaryClass  = {quant-ph}
}

@article{moro2021compiling,
  title   = {Quantum compiling by deep reinforcement learning},
  author  = {Moro, Lorenzo and Paris, Matteo G. A. and Restelli, Marcello and Prati, Enrico},
  journal = {Communications Physics},
  volume  = {4},
  number  = {1},
  pages   = {178},
  year    = {2021}
}

@article{kremer2024transpiling,
  title={Practical and efficient quantum circuit synthesis and transpiling with reinforcement learning},
  author={Kremer, David and Villar, Victor and Paik, Hanhee and Duran, Ivan and Faro, Ismael and Cruz-Benito, Juan},
  journal={arXiv preprint arXiv:2405.13196},
  year={2024}
}

@article{raussendorf2001oneway,
  title   = {A One-Way Quantum Computer},
  author  = {Raussendorf, Robert and Briegel, Hans J.},
  journal = {Physical Review Letters},
  volume  = {86},
  pages   = {5188--5191},
  year    = {2001},
  doi     = {10.1103/PhysRevLett.86.5188}
}

@inproceedings{kingma2015adam,
  title={Adam: A method for stochastic optimization},
  author={Kingma, Diederik P. and Ba, Jimmy},
  booktitle={International Conference on Learning Representations (ICLR)},
  year={2015},
  eprint={1412.6980},
  archivePrefix={arXiv}
}

@article{cerezo2021vqa,
  title   = {Variational quantum algorithms},
  author  = {Cerezo, M. and Arrasmith, Andrew and Babbush, Ryan and Benjamin, Simon C. and Endo, Suguru and Fujii, Keisuke and McClean, Jarrod R. and Mitarai, Kosuke and Yuan, Xiao and Cincio, Lukasz and Coles, Patrick J.},
  journal = {Nature Reviews Physics},
  volume  = {3},
  pages   = {625--644},
  year    = {2021},
  doi     = {10.1038/s42254-021-00348-9}
}

@article{kandala2017hardware,
  title   = {Hardware-efficient variational quantum eigensolver for small molecules and quantum magnets},
  author  = {Kandala, Abhinav and Mezzacapo, Antonio and Temme, Kristan and Takita, Maika and Brink, Markus and Chow, Jerry M. and Gambetta, Jay M.},
  journal = {Nature},
  volume  = {549},
  pages   = {242--246},
  year    = {2017},
  doi     = {10.1038/nature23879}
}

@article{peruzzo2014vqe,
  title   = {A variational eigenvalue solver on a photonic quantum processor},
  author  = {Peruzzo, Alberto and McClean, Jarrod and Shadbolt, Peter and Yung, Man-Hong and Zhou, Xiao-Qi and Love, Peter J. and Aspuru-Guzik, Al{\'a}n and O'Brien, Jeremy L.},
  journal = {Nature Communications},
  volume  = {5},
  pages   = {4213},
  year    = {2014},
  doi     = {10.1038/ncomms5213}
}

@inproceedings{martyniuk2024qassurvey,
  title     = {Quantum Architecture Search: A Survey},
  author    = {Martyniuk, Darya and Jung, Johannes and Paschke, Adrian},
  booktitle = {IEEE International Conference on Quantum Computing and Engineering (QCE)},
  pages     = {1695--1706},
  year      = {2024},
  doi       = {10.1109/QCE60285.2024.00198},
  eprint    = {2406.06210},
  archivePrefix = {arXiv}
}

@article{grimsley2019adapt,
  title   = {An adaptive variational algorithm for exact molecular simulations on a quantum computer},
  author  = {Grimsley, Harper R. and Economou, Sophia E. and Barnes, Edwin and Mayhall, Nicholas J.},
  journal = {Nature Communications},
  volume  = {10},
  number  = {1},
  pages   = {3007},
  year    = {2019}
}

@article{tang2021qubitadapt,
  title   = {qubit-{ADAPT-VQE}: An adaptive algorithm for constructing hardware-efficient ans{\"a}tze on a quantum processor},
  author  = {Tang, Ho Lun and Shkolnikov, V. O. and Barron, George S. and Grimsley, Harper R. and Mayhall, Nicholas J. and Barnes, Edwin and Economou, Sophia E.},
  journal = {PRX Quantum},
  volume  = {2},
  pages   = {020310},
  year    = {2021}
}

@article{zhang2022dqas,
  title   = {Differentiable quantum architecture search},
  author  = {Zhang, Shi-Xin and Hsieh, Chang-Yu and Zhang, Shengyu and Yao, Hong},
  journal = {Quantum Science and Technology},
  volume  = {7},
  number  = {4},
  pages   = {045023},
  year    = {2022}
}

@inproceedings{wu2023quantumdarts,
  title     = {{QuantumDARTS}: Differentiable quantum architecture search for variational quantum algorithms},
  author    = {Wu, Wenjie and Yan, Ge and Lu, Xudong and Pan, Kaisen and Yan, Junchi},
  booktitle = {Proceedings of the 40th International Conference on Machine Learning (ICML)},
  pages     = {37745--37764},
  year      = {2023}
}

@inproceedings{ding2022eqas,
  title     = {Evolutionary quantum architecture search for parametrized quantum circuits},
  author    = {Ding, Li and Spector, Lee},
  booktitle = {Proceedings of the Genetic and Evolutionary Computation Conference Companion (GECCO)},
  pages     = {2190--2195},
  doi       = {10.1145/3520304.3534012},
  year      = {2022}
}

@inproceedings{nicoli2023bayesian,
  title     = {Physics-informed {Bayesian} optimization of variational quantum circuits},
  author    = {Nicoli, Kim A. and Anders, Christopher J. and Funcke, Lena and Hartung, Tobias and Jansen, Karl and K{\"u}hn, Stefan and M{\"u}ller, Klaus-Robert and Stornati, Paolo and Kessel, Pan and Nakajima, Shinichi},
  booktitle = {Advances in Neural Information Processing Systems},
  volume    = {36},
  pages     = {18341--18376},
  year      = {2023}
}

@article{wang2023mcts,
  title   = {Automated quantum circuit design with nested {Monte Carlo} tree search},
  author  = {Wang, Peiyong and Usman, Muhammad and Parampalli, Udaya and Hollenberg, Lloyd C. L. and Myers, Casey R.},
  journal = {IEEE Transactions on Quantum Engineering},
  volume  = {4},
  pages   = {1--20},
  year    = {2023}
}

@article{du2022quantumsupernet,
  title   = {Quantum circuit architecture search for variational quantum algorithms},
  author  = {Du, Yuxuan and Huang, Tao and You, Shan and Hsieh, Min-Hsiu and Tao, Dacheng},
  journal = {npj Quantum Information},
  volume  = {8},
  number  = {1},
  pages   = {62},
  year    = {2022}
}

@article{zhang2021neuralpredictor,
  title   = {Neural predictor based quantum architecture search},
  author  = {Zhang, Shi-Xin and Hsieh, Chang-Yu and Zhang, Shengyu and Yao, Hong},
  journal = {Machine Learning: Science and Technology},
  volume  = {2},
  number  = {4},
  pages   = {045027},
  year    = {2021}
}

@inproceedings{wang2022quantumnas,
  title     = {{QuantumNAS}: Noise-adaptive search for robust quantum circuits},
  author    = {Wang, Hanrui and Ding, Yongshan and Gu, Jiaqi and Lin, Yujun and Pan, David Z. and Chong, Frederic T. and Han, Song},
  booktitle = {IEEE International Symposium on High-Performance Computer Architecture (HPCA)},
  pages     = {692--708},
  doi       = {10.1109/HPCA53966.2022.00057},
  year      = {2022}
}

@inproceedings{he2024trainingfree,
  title     = {Training-free quantum architecture search},
  author    = {He, Zhimin and Deng, Maijie and Zheng, Shenggen and Li, Lvzhou and Situ, Haozhen},
  booktitle = {Proceedings of the AAAI Conference on Artificial Intelligence},
  volume    = {38},
  number    = {11},
  pages     = {12430--12438},
  doi       = {10.1609/aaai.v38i11.29135},
  year      = {2024}
}

@article{sun2026qasurl,
  title   = {Quantum Architecture Search with Unsupervised Representation Learning},
  author  = {Sun, Yize and Wu, Zixin and Tresp, Volker and Ma, Yunpu},
  journal = {Quantum},
  volume  = {10},
  pages   = {1994},
  year    = {2026},
  doi     = {10.22331/q-2026-02-03-1994},
  eprint  = {2401.11576},
  archivePrefix = {arXiv}
}

@article{he2023gsqas,
  title   = {{GSQAS}: Graph Self-supervised Quantum Architecture Search},
  author  = {He, Zhimin and Deng, Maijie and Zheng, Shenggen and Li, Lvzhou and Situ, Haozhen},
  journal = {Physica A: Statistical Mechanics and its Applications},
  volume  = {630},
  pages   = {129286},
  year    = {2023},
  doi     = {10.1016/j.physa.2023.129286},
  eprint  = {2303.12381},
  archivePrefix = {arXiv}
}

@inproceedings{yan2020arch2vec,
  title     = {Does Unsupervised Architecture Representation Learning Help Neural Architecture Search?},
  author    = {Yan, Shen and Zheng, Yu and Ao, Wei and Zeng, Xiao and Zhang, Mi},
  booktitle = {Advances in Neural Information Processing Systems},
  volume    = {33},
  year      = {2020},
  eprint    = {2006.06936},
  archivePrefix = {arXiv}
}

@inproceedings{zhang2019dvae,
  title     = {{D-VAE}: A Variational Autoencoder for Directed Acyclic Graphs},
  author    = {Zhang, Muhan and Jiang, Shali and Cui, Zhicheng and Garnett, Roman and Chen, Yixin},
  booktitle = {Advances in Neural Information Processing Systems},
  volume    = {32},
  pages     = {1588--1600},
  year      = {2019}
}

@article{luo2018nao,
  title={Neural architecture optimization},
  author={Luo, Renqian and Tian, Fei and Qin, Tao and Chen, Enhong and Liu, Tie-Yan},
  journal={Advances in neural information processing systems},
  volume={31},
  year={2018}
}

@inproceedings{tripp2020weighted,
  title     = {Sample-Efficient Optimization in the Latent Space of Deep Generative Models via Weighted Retraining},
  author    = {Tripp, Austin and Daxberger, Erik and Hern{\'a}ndez-Lobato, Jos{\'e} Miguel},
  booktitle = {Advances in Neural Information Processing Systems},
  volume    = {33},
  pages     = {11259--11272},
  year      = {2020}
}

@inproceedings{zhou2020plas,
  title     = {{PLAS}: Latent Action Space for Offline Reinforcement Learning},
  author    = {Zhou, Wenxuan and Bajracharya, Sujay and Held, David},
  booktitle = {Proceedings of the 2020 Conference on Robot Learning (CoRL)},
  series    = {PMLR},
  volume    = {155},
  pages     = {1719--1735},
  year      = {2021}
}

@inproceedings{kingma2014vae,
  title     = {Auto-Encoding Variational {Bayes}},
  author    = {Kingma, Diederik P. and Welling, Max},
  booktitle = {International Conference on Learning Representations (ICLR)},
  year      = {2014},
  eprint    = {1312.6114},
  archivePrefix = {arXiv}
}

@inproceedings{rezende2014stochastic,
  title     = {Stochastic Backpropagation and Approximate Inference in Deep Generative Models},
  author    = {Rezende, Danilo Jimenez and Mohamed, Shakir and Wierstra, Daan},
  booktitle = {Proceedings of the 31st International Conference on Machine Learning (ICML)},
  pages     = {1278--1286},
  year      = {2014}
}

@inproceedings{higgins2017betavae,
  title     = {beta-{VAE}: Learning Basic Visual Concepts with a Constrained Variational Framework},
  author    = {Higgins, Irina and Matthey, Loic and Pal, Arka and Burgess, Christopher and Glorot, Xavier and Botvinick, Matthew and Mohamed, Shakir and Lerchner, Alexander},
  booktitle = {International Conference on Learning Representations (ICLR)},
  year      = {2017}
}

@inproceedings{bowman2016generating,
  title     = {Generating Sentences from a Continuous Space},
  author    = {Bowman, Samuel R. and Vilnis, Luke and Vinyals, Oriol and Dai, Andrew M. and Jozefowicz, Rafal and Bengio, Samy},
  booktitle = {Proceedings of the 20th SIGNLL Conference on Computational Natural Language Learning (CoNLL)},
  pages     = {10--21},
  year      = {2016}
}

@inproceedings{cho2014gru,
  title     = {Learning Phrase Representations using {RNN} Encoder--Decoder for Statistical Machine Translation},
  author    = {Cho, Kyunghyun and van Merri{\"e}nboer, Bart and Gulcehre, Caglar and Bahdanau, Dzmitry and Bougares, Fethi and Schwenk, Holger and Bengio, Yoshua},
  booktitle = {Proceedings of the 2014 Conference on Empirical Methods in Natural Language Processing (EMNLP)},
  pages     = {1724--1734},
  year      = {2014}
}

@article{gomezbombarelli2018automatic,
  title   = {Automatic Chemical Design Using a Data-Driven Continuous Representation of Molecules},
  author  = {G{\'o}mez-Bombarelli, Rafael and Wei, Jennifer N. and Duvenaud, David and Hern{\'a}ndez-Lobato, Jos{\'e} Miguel and S{\'a}nchez-Lengeling, Benjam{\'i}n and Sheberla, Dennis and Aguilera-Iparraguirre, Jorge and Hirzel, Timothy D. and Adams, Ryan P. and Aspuru-Guzik, Al{\'a}n},
  journal = {ACS Central Science},
  volume  = {4},
  number  = {2},
  pages   = {268--276},
  year    = {2018},
  doi     = {10.1021/acscentsci.7b00572}
}

@book{sutton2018rl,
  title     = {Reinforcement Learning: An Introduction},
  author    = {Sutton, Richard S. and Barto, Andrew G.},
  edition   = {2nd},
  publisher = {MIT Press},
  year      = {2018}
}

@article{watkins1992qlearning,
  title   = {Q-learning},
  author  = {Watkins, Christopher J. C. H. and Dayan, Peter},
  journal = {Machine Learning},
  volume  = {8},
  pages   = {279--292},
  year    = {1992},
  doi     = {10.1007/BF00992698}
}

@article{lin1992replay,
  title   = {Self-improving reactive agents based on reinforcement learning, planning and teaching},
  author  = {Lin, Long-Ji},
  journal = {Machine Learning},
  volume  = {8},
  pages   = {293--321},
  year    = {1992},
  doi     = {10.1007/BF00992699}
}

@article{mnih2015dqn,
  title   = {Human-level control through deep reinforcement learning},
  author  = {Mnih, Volodymyr and Kavukcuoglu, Koray and Silver, David and Rusu, Andrei A. and Veness, Joel and Bellemare, Marc G. and Graves, Alex and Riedmiller, Martin and Fidjeland, Andreas K. and Ostrovski, Georg and others},
  journal = {Nature},
  volume  = {518},
  pages   = {529--533},
  year    = {2015},
  doi     = {10.1038/nature14236}
}

@article{vanhasselt2010double,
  title={Double {Q}-learning},
  author={van Hasselt, Hado},
  journal={Advances in Neural Information Processing Systems},
  volume={23},
  pages={2613--2621},
  year={2010}
}

@inproceedings{vanhasselt2016ddqn,
  title     = {Deep Reinforcement Learning with Double {Q}-Learning},
  author    = {van Hasselt, Hado and Guez, Arthur and Silver, David},
  booktitle = {Proceedings of the AAAI Conference on Artificial Intelligence},
  volume    = {30},
  number    = {1},
  pages     = {2094--2100},
  year      = {2016},
  doi       = {10.1609/aaai.v30i1.10295}
}

@inproceedings{kingma2016iaf,
  title     = {Improved Variational Inference with Inverse Autoregressive Flow},
  author    = {Kingma, Diederik P. and Salimans, Tim and Jozefowicz, Rafal and Chen, Xi and Sutskever, Ilya and Welling, Max},
  booktitle = {Advances in Neural Information Processing Systems},
  volume    = {29},
  pages     = {4743--4751},
  year      = {2016}
}

@inproceedings{burda2016iwae,
  title={Importance weighted autoencoders},
  author={Burda, Yuri and Grosse, Roger and Salakhutdinov, Ruslan},
  booktitle={International Conference on Learning Representations (ICLR)},
  year={2016},
  eprint={1509.00519},
  archivePrefix={arXiv}
}

@article{iten2020scinet,
  title     = {Discovering Physical Concepts with Neural Networks},
  author    = {Iten, Raban and Metger, Tony and Wilming, Henrik and del Rio, L{\'\i}dia and Renner, Renato},
  journal   = {Physical Review Letters},
  volume    = {124},
  number    = {1},
  pages     = {010508},
  year      = {2020},
  doi       = {10.1103/PhysRevLett.124.010508},
  eprint    = {1807.10300},
  archivePrefix = {arXiv}
}

@article{taylor2009transfer,
  title   = {Transfer Learning for Reinforcement Learning Domains: A Survey},
  author  = {Taylor, Matthew E. and Stone, Peter},
  journal = {Journal of Machine Learning Research},
  volume  = {10},
  pages   = {1633--1685},
  year    = {2009}
}

@article{zhu2023transfer,
  title   = {Transfer Learning in Deep Reinforcement Learning: A Survey},
  author  = {Zhu, Zhuangdi and Lin, Kaixiang and Jain, Anil K. and Zhou, Jiayu},
  journal = {IEEE Transactions on Pattern Analysis and Machine Intelligence},
  volume  = {45},
  number  = {11},
  pages   = {13344--13362},
  year    = {2023},
  doi     = {10.1109/TPAMI.2023.3292075}
}

@inproceedings{wei2018l2t,
  title     = {Transfer Learning via Learning to Transfer},
  author    = {Wei, Ying and Zhang, Yu and Huang, Junzhou and Yang, Qiang},
  booktitle = {Proceedings of the 35th International Conference on Machine Learning (ICML)},
  series    = {Proceedings of Machine Learning Research},
  volume    = {80},
  pages     = {5085--5094},
  year      = {2018}
}

@inproceedings{vinyals2016order,
  title={Order matters: Sequence to sequence for sets},
  author={Vinyals, Oriol and Bengio, Samy and Kudlur, Manjunath},
  booktitle={International Conference on Learning Representations (ICLR)},
  year={2016},
  eprint={1511.06391},
  archivePrefix={arXiv}
}

@misc{achiam2018spinningup,
  author       = {Achiam, Joshua},
  title        = {Spinning Up as a Deep {RL} Researcher},
  howpublished = {Spinning Up in Deep Reinforcement Learning, OpenAI},
  year         = {2018},
  note         = {\url{https://spinningup.openai.com/en/latest/spinningup/spinningup.html}}
}

@article{kundu2024vqsd,
  title   = {Enhancing variational quantum state diagonalization using reinforcement learning techniques},
  author  = {Kundu, Akash and Bede{\l}ek, Przemys{\l}aw and Ostaszewski, Mateusz and Danaci, Onur and Patel, Yash J. and Dunjko, Vedran and Miszczak, Jaros{\l}aw A.},
  journal = {New Journal of Physics},
  volume  = {26},
  number  = {1},
  pages   = {013034},
  year    = {2024},
  doi     = {10.1088/1367-2630/ad1b7f},
  eprint  = {2306.11086},
  archivePrefix = {arXiv}
}

@misc{javadiabhari2024qiskit,
  title  = {Quantum computing with {Qiskit}},
  author = {Javadi-Abhari, Ali and Treinish, Matthew and Krsulich, Kevin and Wood, Christopher J. and Lishman, Jake and Gacon, Julien and Martiel, Simon and Nation, Paul D. and Bishop, Lev S. and Cross, Andrew W. and Johnson, Blake R. and Gambetta, Jay M.},
  year   = {2024},
  eprint = {2405.08810},
  archivePrefix = {arXiv},
  primaryClass  = {quant-ph}
}

@article{aspuruguzik2005molecular,
  title   = {Simulated quantum computation of molecular energies},
  author  = {Aspuru-Guzik, Al{\'a}n and Dutoi, Anthony D. and Love, Peter J. and Head-Gordon, Martin},
  journal = {Science},
  volume  = {309},
  number  = {5741},
  pages   = {1704--1707},
  year    = {2005},
  doi     = {10.1126/science.1113479}
}

@article{kassal2011chemistry,
  title   = {Simulating chemistry using quantum computers},
  author  = {Kassal, Ivan and Whitfield, James D. and Perdomo-Ortiz, Alejandro and Yung, Man-Hong and Aspuru-Guzik, Al{\'a}n},
  journal = {Annual Review of Physical Chemistry},
  volume  = {62},
  pages   = {185--207},
  year    = {2011},
  doi     = {10.1146/annurev-physchem-032210-103512}
}

@article{omalley2016scalable,
  title   = {Scalable quantum simulation of molecular energies},
  author  = {O'Malley, P. J. J. and Babbush, R. and Kivlichan, I. D. and Romero, J. and McClean, J. R. and Barends, R. and Kelly, J. and Roushan, P. and Tranter, A. and Ding, N. and others},
  journal = {Physical Review X},
  volume  = {6},
  number  = {3},
  pages   = {031007},
  year    = {2016},
  doi     = {10.1103/PhysRevX.6.031007}
}

@article{cao2019chemistry,
  title   = {Quantum chemistry in the age of quantum computing},
  author  = {Cao, Yudong and Romero, Jonathan and Olson, Jonathan P. and Degroote, Matthias and Johnson, Peter D. and Kieferov{\'a}, M{\'a}ria and Kivlichan, Ian D. and Menke, Tim and Peropadre, Borja and Sawaya, Nicolas P. D. and others},
  journal = {Chemical Reviews},
  volume  = {119},
  number  = {19},
  pages   = {10856--10915},
  year    = {2019},
  doi     = {10.1021/acs.chemrev.8b00803}
}

@article{mcardle2020chemistry,
  title   = {Quantum computational chemistry},
  author  = {McArdle, Sam and Endo, Suguru and Aspuru-Guzik, Al{\'a}n and Benjamin, Simon C. and Yuan, Xiao},
  journal = {Reviews of Modern Physics},
  volume  = {92},
  number  = {1},
  pages   = {015003},
  year    = {2020},
  doi     = {10.1103/RevModPhys.92.015003}
}

@article{bauer2020materials,
  title   = {Quantum algorithms for quantum chemistry and quantum materials science},
  author  = {Bauer, Bela and Bravyi, Sergey and Motta, Mario and Chan, Garnet Kin-Lic},
  journal = {Chemical Reviews},
  volume  = {120},
  number  = {22},
  pages   = {12685--12717},
  year    = {2020},
  doi     = {10.1021/acs.chemrev.9b00829}
}

@article{bouland2019rcs,
  title   = {On the complexity and verification of quantum random circuit sampling},
  author  = {Bouland, Adam and Fefferman, Bill and Nirkhe, Chinmay and Vazirani, Umesh},
  journal = {Nature Physics},
  volume  = {15},
  number  = {2},
  pages   = {159--163},
  year    = {2019},
  doi     = {10.1038/s41567-018-0318-2}
}

@article{arute2019supremacy,
  title   = {Quantum supremacy using a programmable superconducting processor},
  author  = {Arute, Frank and Arya, Kunal and Babbush, Ryan and Bacon, Dave and Bardin, Joseph C. and Barends, Rami and Biswas, Rupak and Boixo, Sergio and Brandao, Fernando G. S. L. and Buell, David A. and others},
  journal = {Nature},
  volume  = {574},
  number  = {7779},
  pages   = {505--510},
  year    = {2019},
  doi     = {10.1038/s41586-019-1666-5}
}

@article{madsen2022advantage,
  title   = {Quantum computational advantage with a programmable photonic processor},
  author  = {Madsen, Lars S. and Laudenbach, Fabian and Askarani, Mohsen Falamarzi and Rortais, Fabien and Vincent, Trevor and Bulmer, Jacob F. F. and Miatto, Filippo M. and Neuhaus, Leonhard and Helt, Lukas G. and Collins, Matthew J. and others},
  journal = {Nature},
  volume  = {606},
  number  = {7912},
  pages   = {75--81},
  year    = {2022},
  doi     = {10.1038/s41586-022-04725-x}
}

@article{plesch2011stateprep,
  title   = {Quantum-state preparation with universal gate decompositions},
  author  = {Plesch, Martin and Brukner, {\v{C}}aslav},
  journal = {Physical Review A},
  volume  = {83},
  number  = {3},
  pages   = {032302},
  year    = {2011},
  doi     = {10.1103/PhysRevA.83.032302}
}

@article{cerveralierta2019ame,
  title   = {Quantum circuits for maximally entangled states},
  author  = {Cervera-Lierta, Alba and Latorre, Jos{\'e} Ignacio and Goyeneche, Dardo},
  journal = {Physical Review A},
  volume  = {100},
  number  = {2},
  pages   = {022342},
  year    = {2019},
  doi     = {10.1103/PhysRevA.100.022342}
}

@article{leone2024hea,
  title   = {On the practical usefulness of the Hardware Efficient Ansatz},
  author  = {Leone, Lorenzo and Oliviero, Salvatore F. E. and Cincio, Lukasz and Cerezo, M.},
  journal = {Quantum},
  volume  = {8},
  pages   = {1395},
  year    = {2024},
  doi     = {10.22331/q-2024-07-03-1395}
}

@article{romero2019ucc,
  title   = {Strategies for quantum computing molecular energies using the unitary coupled cluster ansatz},
  author  = {Romero, Jonathan and Babbush, Ryan and McClean, Jarrod R. and Hempel, Cornelius and Love, Peter J. and Aspuru-Guzik, Al{\'a}n},
  journal = {Quantum Science and Technology},
  volume  = {4},
  number  = {1},
  pages   = {014008},
  year    = {2019},
  doi     = {10.1088/2058-9565/aad3e4}
}

@article{anand2022ucc,
  title   = {A quantum computing view on unitary coupled cluster theory},
  author  = {Anand, Abhinav and Schleich, Philipp and Alperin-Lea, Sumner and Jensen, Phillip W. K. and Sim, Sukin and D{\'i}az-Tinoco, Manuel and Kottmann, Jakob S. and Degroote, Matthias and Izmaylov, Artur F. and Aspuru-Guzik, Al{\'a}n},
  journal = {Chemical Society Reviews},
  volume  = {51},
  number  = {5},
  pages   = {1659--1684},
  year    = {2022},
  doi     = {10.1039/D1CS00932J}
}

@article{jordan1928paulische,
  title   = {{\"U}ber das {P}aulische {\"A}quivalenzverbot},
  author  = {Jordan, Pascual and Wigner, Eugene},
  journal = {Zeitschrift f{\"u}r Physik},
  volume  = {47},
  number  = {9},
  pages   = {631--651},
  year    = {1928},
  doi     = {10.1007/BF01331938}
}

@article{bravyi2002fermionic,
  title   = {Fermionic quantum computation},
  author  = {Bravyi, Sergey B. and Kitaev, Alexei Yu.},
  journal = {Annals of Physics},
  volume  = {298},
  number  = {1},
  pages   = {210--226},
  year    = {2002},
  doi     = {10.1006/aphy.2002.6254}
}

@article{seeley2012bk,
  title   = {The {B}ravyi-{K}itaev transformation for quantum computation of electronic structure},
  author  = {Seeley, Jacob T. and Richard, Martin J. and Love, Peter J.},
  journal = {The Journal of Chemical Physics},
  volume  = {137},
  number  = {22},
  pages   = {224109},
  year    = {2012},
  doi     = {10.1063/1.4768229}
}

@inproceedings{zoph2017nas,
  title={Neural architecture search with reinforcement learning},
  author={Zoph, Barret and Le, Quoc V.},
  booktitle={International Conference on Learning Representations (ICLR)},
  year={2017},
  eprint={1611.01578},
  archivePrefix={arXiv}
}

@article{elsken2019nas,
  title   = {Neural architecture search: A survey},
  author  = {Elsken, Thomas and Metzen, Jan Hendrik and Hutter, Frank},
  journal = {Journal of Machine Learning Research},
  volume  = {20},
  number  = {55},
  pages   = {1--21},
  year    = {2019}
}

@article{kumar2025rhodarts,
  title={{RhoDARTS}: Differentiable Quantum Architecture Search with Density Matrix Simulations},
  author={Kumar, Swagat and Zaech, Jan-Nico and Wilmott, Colin Michael and Van Gool, Luc},
  journal={arXiv preprint arXiv:2506.03697},
  year={2025}
}

@article{moflic2025qaser,
  title={{QASER}: Breaking the depth vs. accuracy trade-off for quantum architecture search},
  author={Moflic, Ioana and Paler, Alexandru and Kundu, Akash},
  journal={arXiv preprint arXiv:2511.16272},
  year={2025}
}

@article{kingma2019introduction,
  title   = {An introduction to variational autoencoders},
  author  = {Kingma, Diederik P. and Welling, Max},
  journal = {Foundations and Trends in Machine Learning},
  volume  = {12},
  number  = {4},
  pages   = {307--392},
  year    = {2019},
  doi     = {10.1561/2200000056}
}

\appendix

\section{Deep Q-networks and Double DQN}
\label{sec:prelim-ddqn}

Q-learning~\cite{watkins1992qlearning} turns Eq.~\eqref{eq:bellman} into an update rule by bootstrapping from observed transitions $(s,a,r,s')$. In a deep Q-network (DQN)~\cite{mnih2015dqn} the action values are represented by a neural network $Q_\vartheta(s,\cdot)$ with one output per discrete action, and the network is trained to regress onto the bootstrap target
\begin{equation}
    y^{\mathrm{DQN}}=r+\gamma\,(1-d)\,\max_{a'}Q_{\vartheta^-}(s',a'),
    \label{eq:dqn-target}
\end{equation}
where $d\in\{0,1\}$ marks the end of an episode. Two ingredients stabilise training. First, transitions are stored in a \emph{replay buffer} $\mathcal{D}$ and sampled in random minibatches, which breaks the temporal correlation of consecutive transitions and reuses past experience~\cite{lin1992replay}. Recent work in RL-based QAS shows that how the replay buffer is designed, in particular which transitions are sampled and how often, strongly affects sample efficiency~\cite{kundu2026replay}. Second, the target is computed with a separate \emph{target network} $Q_{\vartheta^-}$ whose parameters are copied from $\vartheta$ only every $K$ updates. The loss is
\begin{equation}
    \mathcal{L}(\vartheta)=\mathbb{E}_{(s,a,r,s',d)\sim\mathcal{D}}\Bigl[\ell_\delta\bigl(y-Q_\vartheta(s,a)\bigr)\Bigr],
    \label{eq:dqn-loss}
\end{equation}
with $\ell_\delta$ the Huber loss, which is quadratic for small errors and linear for large ones and so is less sensitive to outlying targets than the squared error.

\paragraph{Double DQN.}
Because the same (noisy) estimate both selects and evaluates the maximising action in Eq.~\eqref{eq:dqn-target}, $\mathbb{E}[\max_{a'}Q]\ge\max_{a'}\mathbb{E}[Q]$ and DQN systematically overestimates action values~\cite{vanhasselt2010double}. Double DQN (DDQN)~\cite{vanhasselt2016ddqn} decouples the two roles: the online network chooses the next action and the target network evaluates it,
\begin{equation}
    y^{\mathrm{DDQN}}=r+\gamma\,(1-d)\,Q_{\vartheta^-}\Bigl(s',\,\arg\max_{a'}Q_{\vartheta}(s',a')\Bigr).
    \label{eq:ddqn-target}
\end{equation}
This change costs nothing extra, since both networks already exist in DQN, and markedly reduces overestimation. Our implementation follows ref.~\cite{kundu2024kanqas}, which builds on previous work such as ref.~\cite{patel2024crlqas}. $Q_\vartheta$ is a fully connected network trained with the ADAM optimiser~\cite{kingma2015adam} on Eq.~\eqref{eq:dqn-loss}, using the target of Eq.~\eqref{eq:ddqn-target}; the target network is a hard copy of the online network, refreshed every $K$ steps; and the discount factor is tied to the episode budget as $\gamma=\gamma_{\mathrm{f}}^{1/L}$, so that the reward at the final step of an episode is discounted by $\gamma_{\mathrm{f}}$ regardless of $L$. The sizes of the hidden layers, learning rate, replay buffer size, target network update period $K$, final discount $\gamma_{\mathrm{f}}$, and $\epsilon$-greedy decay schedule, changing only the minibatch size.

\section{Hyperparameters}
\label{app:hyperparameters}

Table~\ref{tab:hyperparameters} lists the hyperparameters of all experiments. The agent's hyperparameters are those of KANQAS~\cite{kundu2024kanqas} except for the minibatch size, and are identical for the latent and the gate-based agent. Both agents use an MLP with ReLU activations as their Q-network, not a Kolmogorov-Arnold network.

\begin{table}[h!]
    \centering
    \small
    \caption{Hyperparameters.}
    \label{tab:hyperparameters}
    \begin{tabular}{lll}
        \toprule
        Component & Hyperparameter & Value \\
        \midrule
        DDQN agent & network & MLP, ReLU activations \\
         & hidden layers & $[30,30,30]$ \\
         & optimiser, learning rate & ADAM, $5\cdot10^{-4}$ \\
         & minibatch size & $256$ \\
         & replay buffer size & $15\,000$ \\
         & target network update period $K$ & $50$ \\
         & final discount $\gamma_{\mathrm{f}}$ & $0.005$ \\
         & $\epsilon$ decay factor, $\epsilon_{\min}$ & $0.99995$, $0.05$ \\
        \midrule
        Circuit VAE & embedding size, token width & $16$, $64$ \\
         & GRU hidden size, latent dimension $d$ & $64$, $8$ \\
         & $\beta$ schedule & linear $0\to0.25$ over first $40\%$ \\
         & free bits $\lambda$, word dropout & $1$ nat, $0.5$ \\
         & initial training & $80$ epochs, lr $2\cdot10^{-3}$, batch $64$ \\
         & random training circuits & $2000$ \\
        \midrule
        Latent MDP & step size $\delta$, actions & $0.5$, $2d=16$ \\
         & episode length $L$ & $6$ ($n\le4$), $10$ ($n\ge5$) \\
         & gate budget $T_{\max}$ & $6$, $6$, $8$, $15$, $20$ ($n=2,3,4,5,6$) \\
         & acceptance threshold $\xi$ & $0.98$ \\
         & seeded episode starts & probability $0.5$, from best partial solutions \\
        \midrule
        Refresh & block length & $1000$ episodes \\
         & blocks & $10$ ($n\le4$), $15$ ($n\ge5$) \\
         & fine-tuning & $15$ epochs, lr $5\cdot10^{-4}$ \\
         & pool copies & $25$ \\
         & solution copies & $\min\bigl(300,\max(25,\lfloor3000/N_{\mathrm{sol}}\rfloor)\bigr)$ \\
         & greedy evaluation & $500$ episodes, $\epsilon=0$ \\
        \midrule
        Transfer pool & relabelled / grown variants & up to $200$ / $200$ \\
         & growth & $H$ and $\mathrm{CZ}$ to $1$-$4$ qubits per new qubit \\
        \bottomrule
    \end{tabular}
\end{table}

\FloatBarrier
\section{The four Bell states}
\label{app:bell}

The Bell states are the four maximally entangled states of two qubits,
\begin{equation}
    \ket{\Phi^\pm}=\frac{\ket{00}\pm\ket{11}}{\sqrt2},\qquad
    \ket{\Psi^\pm}=\frac{\ket{01}\pm\ket{10}}{\sqrt2}.
    \label{eq:bell-states}
\end{equation}
In each of them the two qubits are perfectly correlated: the outcome of measuring one qubit fixes the outcome of measuring the other, although each qubit taken on its own is completely random. Together the four states form an orthonormal basis of the two-qubit state space.

All four are prepared by almost the same circuit. A Hadamard gate on the first qubit followed by a $\mathrm{CX}$ gate from the first to the second qubit turns $\ket{00}$ into $\ket{\Phi^+}$. One further single-qubit gate gives each of the others: a $Z$ gate on the first qubit gives $\ket{\Phi^-}$, an $X$ gate on the second qubit gives $\ket{\Psi^+}$, and both gates together give $\ket{\Psi^-}$, up to an overall sign that has no physical meaning.

This is why any of the four is accepted at the two-qubit level. A gate that acts on a single qubit can neither create nor destroy entanglement, so the four states are entangled to exactly the same degree and are equally hard to prepare; they are the same resource written in four ways. The task at $n=2$ is therefore to prepare a maximally entangled pair, not one particular Bell state. In practice we compute the fidelity of the decoded circuit with each of the four states of Eq.~\eqref{eq:bell-states} and use the largest, so a circuit counts as a solution if this largest fidelity reaches the acceptance threshold $\xi$. For $n\ge3$ every level has a single target state.

\FloatBarrier
\section{Effect of the step size on a latent move}
\label{app:step-size}

Table~\ref{tab:step-size} shows what a single action does to the decoded circuit, as a function of the step size $\delta$. For the VAE with which each level starts, we decode $300$ points drawn from the prior, apply each of the $16$ moves $z\to z\pm\delta e_k$ to every point, and decode again. We report how often the decoded circuit changes, and, when it does, the edit distance between the old and the new circuit, that is, the smallest number of gates that have to be inserted, deleted or replaced to turn one into the other.

A step of $\delta=0.1$ leaves the circuit unchanged in about half of the moves at $n=5$ and $n=6$; such a move returns the same fidelity and the same reward, and tells the agent nothing. A step of $\delta=2$ changes the circuit in every move but replaces more than half of its gates, so that the agent can no longer refine a circuit that is almost correct. The value used in this work, $\delta=0.5$, changes the circuit in $88$-$94\%$ of the moves at $n=5$ and $n=6$ and edits $4.6$ to $4.7$ of its $9$ to $14$ gates. With $L=6$ moves per episode for $n\le3$ and $L=10$ for $n\ge5$, an episode covers $L\delta=3$ to $5$ standard deviations of the prior along one axis. At $n=2$, where the circuit space is small, only a third of the moves change the circuit even at $\delta=0.5$.

This measurement characterises the move, not the learning. It shows that $\delta=0.5$ lies between steps that are mostly inert and steps that are no longer local, but it does not establish that this value is optimal for the agent, which would require training at other step sizes. 

\begin{table}[h!]
    \centering
    \small
    \caption{\small\textbf{Effect of one latent move on the decoded circuit, for the starting VAE of each level and $300$ prior samples}. Changed: fraction of the $16$ moves after which the decoded circuit differs. Edit: mean edit distance, in gates, of the moves that change the circuit. The last row gives the mean length of the decoded circuits. Bold: the step size used in this work.}
    \label{tab:step-size}
    \begin{tabular}{lrrrrrrrr}
        \toprule
        & \multicolumn{2}{c}{$n=2$} & \multicolumn{2}{c}{$n=3$} & \multicolumn{2}{c}{$n=5$} & \multicolumn{2}{c}{$n=6$} \\
        \cmidrule(lr){2-3}\cmidrule(lr){4-5}\cmidrule(lr){6-7}\cmidrule(lr){8-9}
        $\delta$ & Changed & Edit & Changed & Edit & Changed & Edit & Changed & Edit \\
        \midrule
        $0.1$  & $10.0\%$ & $1.9$ & $28.1\%$ & $1.7$ & $51.9\%$  & $3.3$ & $41.0\%$ & $3.6$ \\
        $0.25$ & $20.6\%$ & $2.0$ & $54.0\%$ & $1.9$ & $79.2\%$  & $3.7$ & $68.0\%$ & $4.1$ \\
        $\mathbf{0.5}$ & $\mathbf{32.1\%}$ & $\mathbf{2.0}$ & $\mathbf{75.2\%}$ & $\mathbf{2.2}$ & $\mathbf{93.5\%}$ & $\mathbf{4.6}$ & $\mathbf{87.6\%}$ & $\mathbf{4.7}$ \\
        $1.0$  & $45.4\%$ & $2.1$ & $91.7\%$ & $2.8$ & $99.0\%$  & $5.8$ & $97.6\%$ & $5.9$ \\
        $2.0$  & $62.3\%$ & $2.2$ & $98.7\%$ & $3.7$ & $100.0\%$ & $7.4$ & $99.8\%$ & $7.6$ \\
        \midrule
        Mean length (gates) & \multicolumn{2}{c}{$2.6$} & \multicolumn{2}{c}{$5.0$} & \multicolumn{2}{c}{$9.2$} & \multicolumn{2}{c}{$14.1$} \\
        \bottomrule
    \end{tabular}
\end{table}

\FloatBarrier
\section{Effect of the solution bonus}
\label{app:bonus}

The reward of Eq.~\eqref{eq:log-reward} pays $50\,F$ for every step on a solution. To see what happens with a different bonus, we repeat the five-qubit level of the ladder with the factor $50$ replaced by $1$, $10$ and $200$, five seeds each. Everything else is unchanged: the runs use the same transfer pool, the same starting VAE and the same seeds as the first five seeds of the ladder, which serve as the reference with a bonus of $50$. Table~\ref{tab:bonus} lists the results.

Bonuses of $10$, $50$ and $200$ cannot be told apart. The target is found in five of five seeds, in the same range of blocks, and the final success rates lie between $86\%$ and $98\%$ in all three cases. The two lowest rates at a bonus of $200$ belong to seeds that first found the target in block $10$ and were still improving when the run ended.

A bonus of $1$ behaves differently. The target is found as often and as early, in four of five seeds and in blocks $3$ to $7$, but it is learned much less reliably: the four seeds end at $30.0\%$, $64.2\%$, $87.4\%$ and $89.2\%$. With this bonus a step on a solution pays $1$, while a step on the best partial solution, which has $F=1/2$, pays $0.8$ at $n=5$. Reaching the target is then hardly more rewarding than staying next to it, and the agent has little reason to prefer the solution.

Whether a seed finds the target at all does not depend on the bonus, since the bonus is only paid once a solution has been reached. Seed $1$, for instance, finds the target with bonuses of $10$ and $200$ but not with $1$ or $50$, which reflects the variation between runs and not an effect of the bonus. We conclude that the bonus must be clearly larger than the reward of a partial solution, but that its exact value is not important.

\begin{table}[h!]
    \centering
    \small
    \caption{\small\textbf{Effect of the solution bonus at $n=5$ (AME5), five seeds per row, $15$ blocks}. The row in bold is the value used in all other experiments, taken from the first five seeds of the ladder. First solving block and final success rate: min-max over the seeds that found the target.}
    \label{tab:bonus}
    \begin{tabular}{rccc}
        \toprule
        Bonus & Seeds solved & First solving block & Final success rate \\
        \midrule
        $1$   & 4/5 & 3-7  & 30.0-89.2\% \\
        $10$  & 5/5 & 4-8  & 91.6-96.8\% \\
        $\mathbf{50}$ & \textbf{5/5} & \textbf{3-9} & \textbf{88.8-98.4\%} \\
        $200$ & 5/5 & 3-10 & 85.8-97.4\% \\
        \bottomrule
    \end{tabular}
\end{table}

\FloatBarrier
\section{Posterior collapse of the six-qubit VAE}
\label{app:collapse}

Table~\ref{tab:collapse} documents the posterior collapse described in Sec.~\ref{sec:method-vae}. Each row is one six-qubit VAE trained without free bits and without word dropout on $400$ transfer-pool circuits, each repeated $25$ times, and $6000$ random circuits; no row uses the target state. The first row is the default configuration, which is sufficient for the two-qubit VAE. Its KL divergence on held-out circuits is $0.002$ nats, that is, the encoder sends every circuit to almost the same point. As a consequence none of the pool circuits is reconstructed, and $1000$ samples of the prior decode to only $132$ distinct circuits, the most frequent of which appears $52$ times. A healthy VAE of the same size decodes the same samples to between $850$ and $1000$ distinct circuits, as Table~\ref{tab:gate-order} shows.

Neither longer training nor a larger model removes the collapse: doubling the number of epochs, the hidden size, or the hidden size and the latent dimension together leaves the KL divergence below $0.01$ nats. Lowering the weight of the KL term to $\beta=0.1$ does keep the latent space in use (KL of $7.5$ nats, $997$ distinct circuits), but at the price of a weaker pull towards the prior, and still without reconstructing a single pool circuit. Free bits remove the collapse at the standard weight $\beta=0.25$, and reconstruction then requires the canonical gate order; both are shown in Table~\ref{tab:gate-order}. As there, each row is a single training run.

\begin{table}[h!]
    \centering
    \small
    \caption{\small\textbf{Six-qubit circuit VAEs trained without free bits and word dropout}. KL: on held-out circuits (nats); reconstruction: exact reconstruction accuracy on the pool circuits; distinct and most frequent: number of distinct circuits among $1000$ decoded prior samples, and how often the most frequent one appears.}
    \label{tab:collapse}
    \begin{tabular}{lrrrr}
        \toprule
        Configuration & KL & Reconstruction & Distinct & Most freq. \\
        \midrule
        Standard ($\beta=0.25$, $40$ epochs, hidden size $64$, $d=8$) & $0.002$ & $0.0\%$ & $132$ & $52$ \\
        $80$ epochs                                   & $0.001$ & $0.0\%$ & $186$ & $74$ \\
        Hidden size $128$                             & $0.004$ & $0.0\%$ & $463$ & $27$ \\
        Hidden size $128$, $d=16$                     & $0.009$ & $0.0\%$ & $513$ & $22$ \\
        $\beta=0.1$                                   & $7.5$   & $0.0\%$ & $997$ & $2$ \\
        \bottomrule
    \end{tabular}
\end{table}

\FloatBarrier
\section{Effect of the gate order on the circuit VAE}
\label{app:gate-order}

Table~\ref{tab:gate-order} reports the analysis behind the canonical gate order of Sec.~\ref{sec:method-encoding}. Each row is one six-qubit VAE ($d=8$) trained on a transfer pool grown from the five-qubit AME solutions of an earlier run, by the procedure of Sec.~\ref{sec:method-transfer}, each pool circuit repeated $25$ times, together with random circuits. We report three properties of the trained model: the KL divergence on held-out circuits, which is close to zero for a collapsed posterior; the \emph{exact reconstruction accuracy}, the fraction of pool circuits that are reproduced gate for gate when encoded to their latent mean and decoded again; and the number of distinct circuits obtained by decoding $1000$ samples of the prior.

With the pool circuits in a random order of their commuting gates, no regularisation makes the pool reconstructible. Word dropout alone leaves the posterior collapsed (KL of $0.003$ nats, $139$ distinct circuits). Free bits remove the collapse, and the prior then decodes to almost $1000$ distinct circuits, but the reconstruction accuracy stays at or below $0.8\%$: the model uses its latent space, yet cannot recover a circuit from its own latent point, because it cannot predict in which of many equivalent orders the gates were written. With a pool built by the same procedure but written in canonical order, and otherwise identical settings, the reconstruction accuracy rises from $0.8\%$ to $79\%$. Training for $80$ instead of $40$ epochs raises it to $87\%$, the configuration used for the six-qubit level; weaker regularisation ($\lambda=0.5$, word dropout $0.25$) lowers it to $47\%$. The gate order, not the regularisation, is therefore what makes the transferred circuits representable.

Two caveats apply. Each row is a single training run. The two pools are separate random draws of the same procedure, grown from the same five-qubit solutions: $400$ circuits in random order and $272$ in canonical order, where variants that differ only in their gate order coincide.

\begin{table}[h!]
    \centering
    \small
    \caption{\small\textbf{Six-qubit circuit VAEs trained on grown transfer pools written in random or in canonical gate order}. $\lambda$: free bits per latent dimension (nats); KL: on held-out circuits (nats); reconstruction: exact reconstruction accuracy, the fraction of pool circuits reproduced gate for gate by encoding and decoding; distinct: distinct circuits among $1000$ decoded prior samples. $^\dagger$With a cyclical $\beta$ schedule of four cycles.}
    \label{tab:gate-order}
    \begin{tabular}{lcccccrr}
        \toprule
        Gate order & $\lambda$ & Word dropout & Random circuits & Epochs & KL & Reconstruction & Distinct \\
        \midrule
        Random    & $0.5$ & $0$    & $6000$ & $40$ & $3.9$   & $0.0\%$  & $949$ \\
        Random    & $0$   & $0.25$ & $6000$ & $40$ & $0.003$ & $0.0\%$  & $139$ \\
        Random    & $0.5$ & $0.25$ & $6000$ & $40$ & $3.7$   & $0.0\%$  & $999$ \\
        Random    & $0.5$ & $0.25^\dagger$ & $6000$ & $40$ & $3.9$ & $0.0\%$ & $998$ \\
        Random    & $1.0$ & $0.5$  & $6000$ & $40$ & $7.7$   & $0.8\%$  & $1000$ \\
        \midrule
        Canonical & $1.0$ & $0.5$  & $6000$ & $40$ & $7.6$   & $79.0\%$ & $925$ \\
        Canonical & $1.0$ & $0.5$  & $2000$ & $40$ & $7.7$   & $75.0\%$ & $854$ \\
        Canonical & $1.0$ & $0.5$  & $2000$ & $80$ & $7.8$   & $87.1\%$ & $848$ \\
        Canonical & $0.5$ & $0.25$ & $2000$ & $80$ & $4.0$   & $47.4\%$ & $549$ \\
        \bottomrule
    \end{tabular}
\end{table}

\FloatBarrier
\section{Details of the noise ablation}
\label{app:noise}

\paragraph{Noise model.} After every gate we apply a depolarising channel to the qubits the gate acts on. For a gate on $k$ qubits, with $k=1$ or $2$, it maps the state $\rho$ to
\begin{equation}
    \rho \;\mapsto\; (1-p_k)\,\rho + p_k\,\mathrm{Tr}_{q}(\rho)\otimes\frac{I}{2^k},
    \label{eq:depolarizing}
\end{equation}
where $\mathrm{Tr}_q$ is the partial trace over these $k$ qubits and $I/2^k$ is their maximally mixed state. We use $p_1=0.001$ and $p_2=0.01$. The circuit is simulated as a density matrix, and the fidelity with the target is $F=\langle\psi|\rho|\psi\rangle$. Our implementation agrees with the density-matrix simulation of Qiskit to numerical precision.

\paragraph{Where the noise enters.} Noise enters wherever a circuit is evaluated: in the reward the agent receives after every move, in the test of whether an episode is solved, and in the selection of the circuits used to refresh the VAE and to build the transfer pool of the next level. The VAE itself is trained on circuits, that is, on sequences of gates, and is not changed. The greedy evaluation uses the same noisy fidelity and threshold.

\paragraph{Acceptance threshold.} Our circuits and targets are stabiliser circuits and stabiliser states, so without noise the fidelity of a circuit with the target is either $0$ or a power of $1/2$: an exact solution has $F=1$ and the best non-solution has $F=1/2$. Under noise, the exact solutions we observe have $F$ between $0.89$ and $0.99$, depending on their number of gates, as listed in Table~\ref{tab:noise-details}. The threshold $\xi=0.75$ lies halfway between the two values. For the bookkeeping of the pool, which ranks circuits by the number of stabilisers they share with the target, a noisy fidelity $F$ is assigned to the largest power $2^{-j}$ with $F\ge0.75\cdot2^{-j}$. On the pools of the noiseless ladder, this assignment recovers the noiseless fidelity of every circuit.

\paragraph{The two settings.} In the first setting, the five-qubit level starts from the transfer pool and the initial VAE of the noiseless ladder, and the six-qubit pool is grown from the solutions of a noisy five-qubit run. In the second setting, the whole ladder $2\to3\to5\to6$ is run with noise, and each pool is grown from a noisy run of the level below. In both, the pool is taken from the seed that solved the most training episodes. No grown circuit solves the next target. All other settings are those of the noiseless ladder.

\paragraph{Solutions.} Table~\ref{tab:noise-details} lists, for each level, when the target was first found and what the solutions look like. All solutions have a noiseless fidelity of $1$. The number of gates varies between runs in the same way as without noise: at $n=5$, some runs settle on circuits of $12$ gates and others on circuits of $15$, in both settings, so the noise does not visibly favour shorter circuits at this strength.

\begin{table}[h!]
    \centering
    \small
    \caption{\small\textbf{Solutions found under depolarising noise, five seeds per row}. First solving block: over the seeds that found the target. Solutions: distinct solution circuits in the final pool of a run. Noisy fidelity and gates: range over all solutions.}
    \label{tab:noise-details}
    \begin{tabular}{llcccc}
        \toprule
        Setting & Target & First solving block & Solutions per run & Noisy fidelity & Gates \\
        \midrule
        Noise at $n=5,6$     & AME5 & 5-6  & 297-537   & 0.924-0.941 & 12-14 \\
                             & AME6 & 1    & 468-582   & 0.907       & 17 \\
        \midrule
        Noise at every level & Bell & 1    & 1817-3031 & 0.956-0.992 & 2-7 \\
                             & GHZ  & 1    & 861-1173  & 0.958-0.983 & 3-6 \\
                             & AME5 & 3-10 & 69-1926   & 0.911-0.941 & 12-15 \\
                             & AME6 & 1    & 398-698   & 0.888-0.920 & 16-19 \\
        \bottomrule
    \end{tabular}
\end{table}

\end{document}